\documentclass[journal=jacsat,manuscript=communication,etalmode=truncate,keywords=true]{achemso}
\setkeys{acs}{articletitle=true,maxauthors=10,etalmode=truncate,keywords=true}
\usepackage{amsmath,amssymb,units,txfonts}
\usepackage{amsfonts}

\usepackage{graphicx}
\usepackage{MnSymbol}
\usepackage{helvet,wasysym}
\usepackage[usenames,dvipsnames]{color}
\usepackage{colortbl,cuted}
\newcommand{\epeak}{\varepsilon_{\textit{peak}}^{\textit{PDOS}}}
\usepackage[utf8x]{inputenc}
\usepackage{xfrac}
\definecolor{JCTCGreen}{RGB}{11,122,64}
\definecolor{abstractcolor}{RGB}{255,255,255}%{255,243,201}
\makeatletter\newenvironment{abstractbox}{%
   \begin{lrbox}{\@tempboxa}\begin{minipage}{0.988\textwidth}}{\end{minipage}\end{lrbox}%
   \colorbox{abstractcolor}{\usebox{\@tempboxa}}
}\makeatother

\renewcommand*\authorfont{\flushleft}%\bf}
\renewcommand*\affilfont{\mdseries\flushleft\textrm}
\renewcommand*\titlesize{\Large}
\renewcommand*\titlefont{\flushleft\sf\bfseries}
\def\refname{\sffamily\color{JCTCGreen}\Large$\blacksquare$\normalsize\ REFERENCES}
\usepackage{titlesec}
\titleformat{\section}{\bfseries\sffamily\color{JCTCGreen}}{\thesection.~}{0pt}{}
\titleformat{\subsection}[runin]{\bfseries\sffamily\normalsize}{\indent\thesubsection.~}{0pt}{}[.]
\titlespacing{\subsection}{0pt}{0pt}{*1}
\titleformat{\subsubsection}{\bfseries\sffamily\normalsize}{\thethesubsection.~}{0pt}{}
\titlespacing{\subsubsection}{0pt}{0pt}{*0}

\newcommand{\new}[1]{{\color{black}{#1}}}
\newcommand{\old}[1]{}

\title{Quasiparticle interfacial level alignment of highly hybridized frontier levels:
H$_{\text{2}}$O on TiO$_{\text{2}}$(110)}
\author{Annapaola Migani}
\email{annapaola.migani@cin2.es}
\affiliation[ICN2]{\footnotemark[2]{\ } ICN2 - Institut Catal\`{a} de Nanoci\`{e}ncia i Nanotecnologia, ICN2 Building, Campus UAB, E-08193 Bellaterra (Barcelona), Spain}
\alsoaffiliation[CSIC]{\newline\footnotemark[3]{\ } CSIC - Consejo Superior de Investigaciones Cient{\'{\i}}ficas, ICN2 Building, Campus UAB, E-08193 Bellaterra (Barcelona), Spain}
\author{Duncan J.~Mowbray}
\affiliation[UPV/EHU]{\newline\footnotemark[5]{\ } Nano-Bio Spectroscopy Group and ETSF Scientific Development Center, Departamento de F{\'{\i}}sica de Materiales, Universidad del Pa{\'{\i}}s Vasco UPV/EHU  and DIPC, E-20018 San Sebasti\'{a}n, Spain}
\author{Jin Zhao}
\affiliation[USTC]{\newline\footnotemark[4]{\ } Department of Physics and ICQD/HFNL, University of Science and Technology of China, Hefei, Anhui 230026, China}
\alsoaffiliation[SICQI]{\newline\footnotemark[6]{\ } Synergetic Innovation Center of Quantum Information \& Quantum Physics, University of Science and Technology of China, Hefei, Anhui 230026, China}
\author{Hrvoje Petek}
\affiliation[UP]{\newline{$^\perp$}Department of Physics and Astronomy, University of Pittsburgh, Pittsburgh, Pennsylvania 15260, USA}
\begin{document}
\maketitle

\begin{strip}
\vspace{-1.cm}

\noindent{\color{JCTCGreen}{\rule{\textwidth}{0.5pt}}}
\begin{abstractbox}
\begin{tabular*}{17cm}{b{7.63cm}r}
\noindent\textbf{\color{JCTCGreen}{ABSTRACT:}}
 Knowledge of the frontier levels' alignment prior to photo-irradiation is necessary to achieve a complete quantitative  description of H$_2$O  photocatalysis on TiO$_2$(110). Although H$_2$O on rutile TiO$_2$(110) has been thoroughly studied both experimentally and theoretically, a quantitative value for the energy of the highest H$_2$O occupied levels is still lacking. For experiment, this is due to the H$_2$O levels being obscured by hybridization with TiO$_2$(110) levels in the difference spectra obtained via ultraviolet  photoemission spectroscopy (UPS).  For theory, this is due to  inherent  difficulties in  properly describing many-body effects  at the H$_2$O--TiO$_2$(110) interface.  Using the projected density
&\includegraphics[width=9cm]{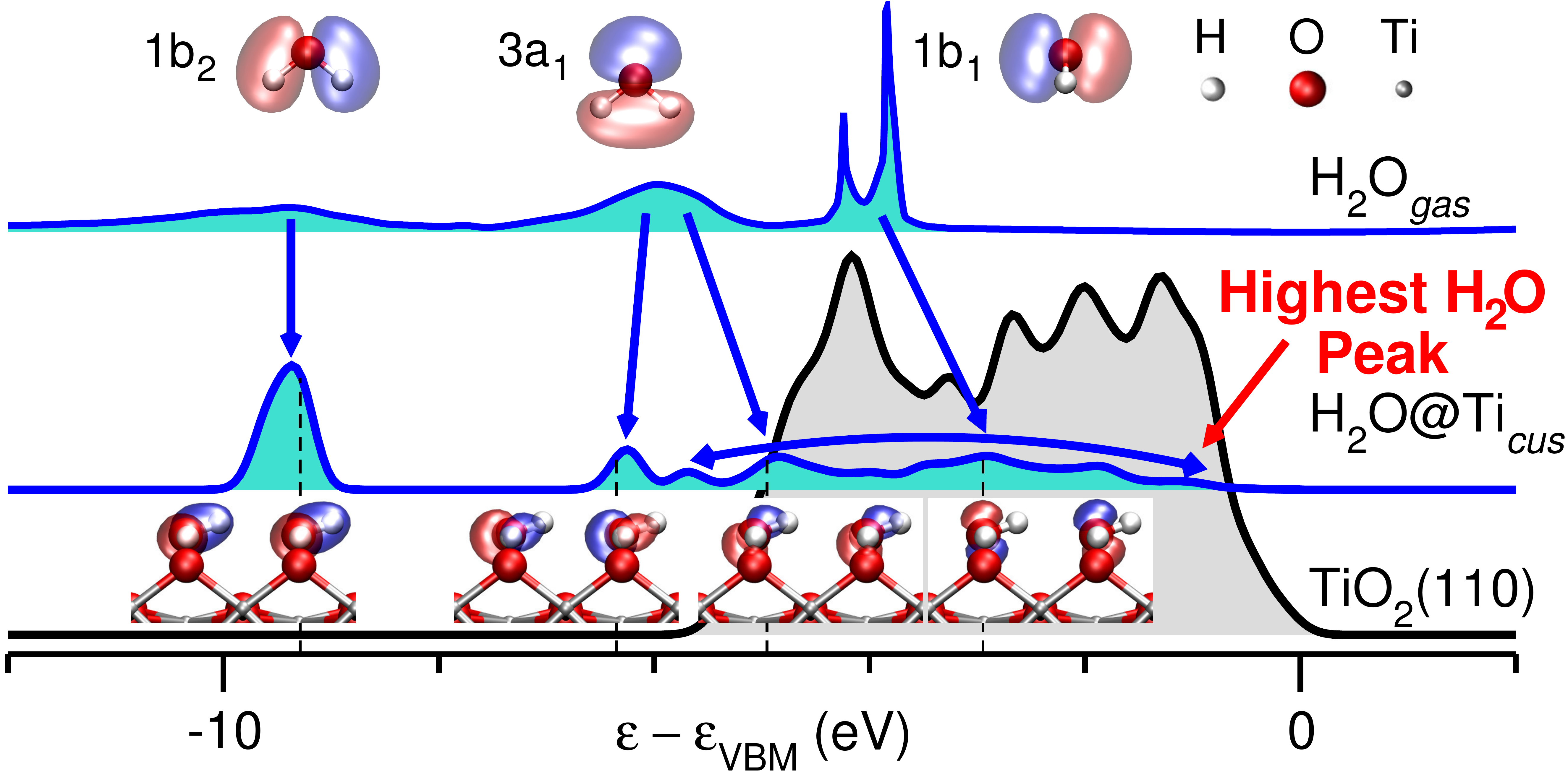}\\
\multicolumn{2}{p{17cm}}{ of states (DOS)  from state-of-the-art  quasiparticle \new{(QP) }$G_0W_0$, we disentangle the adsorbate and surface contributions to the complex UPS spectra of H$_2$O on TiO$_2$(110). We perform this separation as a function of H$_2$O coverage and dissociation on stoichiometric and reduced surfaces.  Due to hybridization with the TiO$_2$(110) surface, the H$_2$O 3a$_1$ and 1b$_1$ levels are broadened into several peaks  between $5 $ and $1$~eV  below the TiO$_2$(110) valence band maximum (VBM). These peaks have both intermolecular and interfacial bonding and antibonding character. We find the highest occupied  levels of H$_2$O adsorbed intact and dissociated on stoichiometric TiO$_2$(110) are  $1.1$ and $0.9$~eV below the VBM.  We also find a similar energy of 1.1~eV for the highest occupied levels of H$_2$O when adsorbed dissociatively on a bridging O vacancy of the reduced surface.  In both cases, these energies are significantly higher (by 0.6 to 2.6~eV) than those \old{derived}\new{estimated} from UPS difference spectra\new{, which are inconclusive in this energy region}. \new{ Finally, we apply self-consistent QP$GW$ (scQP$GW$1) to obtain the ionization potential of the H$_2$O--TiO$_2$(110) interface.}}
\end{tabular*}
\end{abstractbox}
\noindent{\color{JCTCGreen}{\rule{\textwidth}{0.5pt}}}
\end{strip}

\def\bigfirstletter#1#2{{\noindent
    \setbox0\hbox{{\color{JCTCGreen}{\Huge #1}}}\setbox1\hbox{#2}\setbox2\hbox{(}%
    \count0=\ht0\advance\count0 by\dp0\count1\baselineskip
    \advance\count0 by-\ht1\advance\count0 by\ht2
    \dimen1=.5ex\advance\count0 by\dimen1\divide\count0 by\count1
    \advance\count0 by1\dimen0\wd0
    \advance\dimen0 by.25em\dimen1=\ht0\advance\dimen1 by-\ht1
    \global\hangindent\dimen0\global\hangafter-\count0
    \hskip-\dimen0\setbox0\hbox to\dimen0{\raise-\dimen1\box0\hss}%
    \dp0=0in\ht0=0in\box0}#2}

%\bigfirstletter
\section{INTRODUCTION}

The photooxidation activity of a surface is determined by the interfacial level alignment between the occupied adsorbate levels and those of the substrate \cite{YatesChemRev,HendersonSurfSciRep}.
Water photooxidation on TiO$_{\mathrm{2}}$ has attracted enormous attention \cite{WaterDissociationJACS2012Jin,SprikChemCatChem,SprikSchematic,NakamuraJACS2004,ImanishiJACS2007,CuKTanjaJACS2014,SprikAngewandte2014,Tritsaris-KaxirasJPCC2014} for energy applications  \cite{Crabtree,DOE} based on H$_{\mathrm{2}}$ production \cite{FujishimaNature}. This reaction also plays an important role in photocatalytic environmental remediation and surface self-cleaning/sterilizing.\cite{YatesChemRev,HendersonSurfSciRep,FujishimaReview} 
This is because the resulting hydroxyl radicals are the key intermediates in the oxidative degradation of organic species \cite{PSalvador2007,PSalvador2011ProgSurfSci}.   To understand water photooxidation, it is necessary to understand the interfacial level alignment between the occupied levels of H$_{\text{2}}$O and the TiO$_{\mathrm{2}}$ substrate \cite{PrezhdoChemRev2013}. 

Experimentally, the most common approach to access the adsorbate levels is to take the difference between the covered and clean surface spectra from photoemission spectroscopy. However, when the adsorbate and surface levels are strongly hybridized, it becomes difficult to disentangle the adsorbate and surface contributions to the UPS spectra using only the difference spectra\cite{DeSegovia}. For example, shifting of the surface levels due to hybridization or band bending may completely obscure the adsorbate levels \cite{DeSegovia}. Further, the adsorbate levels near the valence band maximum (VBM) are the most likely to be obscured. It is precisely these levels that are most important for photooxidation processes.  Using a theoretical approach, one can directly disentangle the molecular levels by projecting the density of states (DOS) of the interface onto the atomic orbitals of the molecule. 
Altogether, this makes  a  robust theoretical approach necessary to accurately predict the alignment of the adsorbate and substrate levels, and separate the adsorbate and surface spectra. 

A robust theoretical treatment requires quasiparticle (QP) $G_0W_0$  to capture the anisotropic screening of the electron--electron interaction at the interface \cite{RenormalizationLouie,JuanmaRenormalization1,GiustinoPRL2012}. As previously demonstrated for CH$_3$OH on TiO$_2$(110), QP $G_0W_0$ is necessary to obtain even a qualitative description of the level alignment \cite{OurJACS,MiganiLong,MiganiInvited}.  For this interface, the occupied levels of the molecule are only weakly hybridized with the surface levels.  This allowed an unambiguous comparison to the photoemission difference spectrum \cite{OurJACS}. However, for H$_2$O on rutile TiO$_2$(110), this is not the case. 

The occupied molecular levels of H$_{\text{2}}$O on single crystal rutile TiO$_{\text{2}}$(110) have been probed via ultraviolet photoemission spectroscopy (UPS) \cite{DeSegovia,ThorntonH2ODissTiO2110,Krischok} and metastable impact electron spectroscopy (MIES)\cite{Krischok}.  These experiments were performed under ultrahigh vacuum (UHV) conditions from low to room temperature \cite{ThorntonH2ODissTiO2110}, from 0.01 to 100 L H$_2$O exposure \cite{DeSegovia}, and for various surface preparations resulting in either reduced TiO$_{2-x}$(110) with surface oxygen defects or ``nearly-perfect'' TiO$_{2}$(110) \cite{DeSegovia}.  Altogether, these experiments have addressed the long-standing controversy as to where and how H$_2$O adsorbs and dissociates on TiO$_2$(110) \cite{HendersonControversy,LindanControversyPRL,NorskovVacanciesPRL2001,WaterTiO2ControversyPRL2004,LindanMonolayer,MichaelidesDynamics,Walle2013H2ODiss,Walle2014H2ODiss1,Walle2014H2ODiss2,DupuisPRL2009H2ODiss}.

At 150~K the photoemission difference spectrum between H$_2$O covered and clean TiO$_2$(110) surfaces consists of three peaks, which are attributed to intact H$_2$O adsorbed on Ti coordinately unsaturated sites (Ti$_{\textit{cus}}$) \cite{ThorntonH2ODissTiO2110}.  Upon heating to 300 K, the difference spectrum's three-peak structure evolves into a two-peak structure, which is attributed to dissociated H$_2$O adsorbed on bridging O vacancies (O$_{\textit{br}}^{\textit{vac}}$), i.e., O$_{\textit{br}}$H surface species \cite{ThorntonH2ODissTiO2110}.   This assignment of the UPS spectra to intact (I) H$_2$O@Ti$_{\textit{cus}}$ or dissociated (D) H$_2$O@O$_{\textit{br}}^{\textit{vac}}$ is based on the peak energy separations being consistent with  those reported for H$_2$O\cite{WaterSpectra} in gas phase or OH$^-$ in NaOH\cite{NaOH}. 

A comparison to the H$_2$O and OH$^-$ peaks is robust for the molecular levels that lie below and have little hybridization with the surface DOS. However, the adsorbate levels that lie within the surface valence band may significantly hybridize with the surface, with a single molecular level contributing to many interfacial levels. These interfacial levels are thus not easily associated with H$_2$O and OH$^-$ levels. This is exacerbated by the mixing of the molecular levels due to symmetry breaking at the interface.  As a result, ``between 5 and 8 eV'' below the Fermi level, experimentally they ``are unable to produce reliable difference structures'' from the UPS spectra obtained for ``nearly-perfect'' TiO$_{2}$(110) exposed to H$_2$O at 160~K \cite{DeSegovia}.

Using the QP $G_0W_0$ H$_2$O projected DOS (PDOS), we have disentangled the adsorbate and surface contributions to the UPS spectra within this difficult energy range. This has been done as a function of H$_2$O coverage and dissociation on stoichiometric and reduced surfaces.  In so doing, we provide quantitative values for the energies of the highest H$_2$O  occupied levels, prior to photo-irradiation, for a number of experimentally relevant\cite{NakamuraJACS2004,ImanishiJACS2007,PSalvador2007,WaterDissociationJACS2012Jin,SprikSchematic}  H$_2$O--TiO$_2$(110) structures.

\new{To directly compare to red-ox potentials, the important quantities for determining photoelectrocatalytic activity, one needs the alignment relative to the vacuum level, $E_{\textit{vac}}$.\cite{SprikPCCPREview2012,SprikH2OAlignment}  With this, one obtains the ionization potential directly from $-\epeak + E_{\textit{vac}}$.  To obtain a more accurate absolute level alignment, we employ our recently introduced self-consistent QP $GW$\cite{KressescGW,SchilfgaardeQSGW,QSGW} technique scQP$GW1$ \cite{OurJACS}.   }

The presentation of the results is organized as follows. First, we focus on the H$_2$O levels that lie below and have little hybridization with the substrate DOS. This is done for intact  H$_2$O@Ti$_{\textit{cus}}$  in Section~\ref{Sect:Intact} and dissociated H$_2$O@O$_{\textit{br}}^{\textit{vac}}$ in Section~\ref{Sect:Dissociated}.  \new{Further, in Section~\ref{Sect:xc-functionals}, we shown that these results are rather independent of the choice of xc-functional. }In so doing we provide evidence for a \new{robust }semi-quantitative agreement with the UPS difference spectra for the adsorbate levels for which an unambiguous comparison with the experiment is possible. For a more complete understanding of the UPS experiments,  in Section~\ref{Sect:Dependence} we analyze the H$_2$O PDOS for a variety of other H$_2$O structures on the stoichiometric and reduced surfaces. These may form under different experimental  conditions and surface preparations.  \old{Finally,  in}\new{In} Section~\ref{Sect:Alignment} we focus on the highest  H$_2$O occupied levels, which are  significantly hybridized with the substrate DOS. The success of the QP $G_0W_0$ PDOS strategy for the lower-energy part of the UPS difference spectra provides support for our results  in this difficult spectral region, where a straightforward comparison with experiment is  not possible.\new{  Finally, in Section~\ref{Sect:EvacAlignment}, we employ scQP$GW$1 to obtain an improved absolute level alignment relative to $E_{\textit{vac}}$, and thus estimate the ionization potential of the H$_{\text{2}}$O--TiO$_{\text{2}}$(110) interface.}

\section{METHODOLOGY}\label{Sect:Methodology}

Our QP $G_0W_0$ calculations\cite{GW,AngelGWReview,KresseG0W0} have been performed using \textsc{vasp} within the projector augmented wave (PAW) scheme \cite{kresse1999}.   The $G_0W_0$ calculations are based on Kohn-Sham wavefunctions and eigenenergies from density functional theory (DFT) obtained using a generalized gradient approximation (PBE) \cite{PBE} for the exchange correlation (xc)-functional \cite{kresse1996b}. The dependence of the QP $G_0W_0$ DOS and PDOS on the DFT xc-functional has been tested for 1 ML intact H$_2$O@Ti$_{\textit{cus}}$ of stoichiometric TiO$_2$(110)\new{ and \sfrac{1}{2}ML dissociated H$_2$O@O$_{\textit{br}}^{\textit{vac}}$ of defective TiO$_{2-\text{\sfrac{1}{4}}}$(110) with \sfrac{1}{2}ML of O$_{\textit{br}}^{\textit{vac}}$}. For \old{this structure}\new{these structures}, $G_0W_0$ calculations based on the local density approximation (LDA) \cite{LDA}, \old{and }van der Waals (vdW-DF) \cite{vdW-DF}\new{, and the range-separated hybrid (HSE) \cite{HSE} } xc-functionals have been carried out for comparison with the PBE based $G_0W_0$ calculations.  \new{In particular, we use the HSE06~\cite{HSE06} variant of the HSE xc-functional.}

In the QP $G_0W_0$ approach, the contribution to the Kohn-Sham (KS) eigenvalues from the exchange and correlation (xc)-potential $V_{xc}$ is replaced by the self   energy $\Sigma = i G W$, where $G$ is the Green's function and $W$ is the screening \cite{GW} based on the KS wavefunctions \cite{AngelGWReview}.  The dielectric function is obtained from linear response time-dependent (TD) density functional theory (DFT) within the random phase approximation (RPA), including local field effects \cite{KresseG0W0}.  From $G_0W_0$ one obtains first-order QP corrections to the KS eigenvalues, but retains the KS wavefunctions.\new{  Since our aim is to compare the computed interfacial level alignment with measured UPS spectra, it is most consistent to align the QP $G_0W_0$ levels with the VBM.}

\new{We find $E_{\textit{vac}}$, i.e., the effective potential far from the surface, from $G_0W_0$ is essentially the same as the $E_{\textit{vac}}$ from DFT. In other words, the effective potential is unchanged by $G_0W_0$.} 
\new{To obtain a more accurate absolute QP level alignment relative to $E_{\textit{vac}}$, we employ a self-consistent QP $GW$ approach\cite{KressescGW}.  In particular, by employing the scQP$GW$1 approach, we obtain both a QP PDOS comparable to that from QP $G_0W_0$ and an improved alignment relative to $E_{\textit{vac}}$.
%, i.e., the effective potential far above the surface
\cite{OurJACS,MiganiLong}.  Here, 25\%, 25\%, and 50\%, of the QP self energies are ``mixed'' with the DFT xc-potential over three self-consistent QP $GW$ cycles \cite{KressescGW}, respectively. If, instead, 100\% of the DFT xc-potential were replaced by QP self energy in a single self-consistent QP $GW$ cycle, one would exactly obtain the QP $G_0W_0$ eigenvalues\old{, but retain the $E_{\textit{vac}}$ from DFT}. However, this mixing is required to obtain a smooth convergence of both the QP wavefunctions and the absolute QP level alignment.  To fully converge our self-consistent QP $GW$ calculations (scQP$GW$), we perform a further eight cycles, with each introducing a further 25\% of the QP self energy.}

 The geometries have been fully relaxed using LDA\cite{LDA}, PBE\cite{PBE}, or vdW-DF\cite{vdW-DF} xc-functionals, with all forces $\lesssim$ 0.02 eV/\AA.  \new{HSE calculations are performed for the relaxed geometries obtained with PBE. }We employ a plane-wave energy cutoff of 445 eV, an electronic temperature $k_B T\approx0.2$ eV with all energies extrapolated to $T\rightarrow 0$ K, and a PAW pseudopotential for Ti which includes the 3$s^2$ and 3$p^6$ semi-core levels.  All calculations have been performed spin unpolarized.

For the clean stoichiometric TiO$_2$(110) surface \cite{MiganiLong} we have used a four layer slab and an orthorhombic  $1\times1$ unit cell of $6.497 \times 2.958\times 40$ \AA$^3$, i.e., 
\begin{equation}
\left(\begin{array}{ccc}  \sqrt{2}a & 0& 0\\ 0 & c & 0\\ 0 & 0 & \sqrt{2}{a} + D\end{array}\right),
\end{equation}
 where $D \approx 27$~\AA\ is the vacuum thickness and $a$ and $c$ are the experimental lattice parameters for bulk rutile TiO$_{\text{2}}$ ($a=4.5941$ \AA, $c=2.958$ \AA) \cite{TiO2LatticeParameters}.  We have employed a $\Gamma$-centered $4\times 8\times1$ \textbf{k}-point mesh, and 320 bands = 9\sfrac{1}{3} unoccupied bands per atom, i.e.\ including all levels up to 26~eV above the valence band maximum (VBM).

For the clean reduced TiO$_{2-\sfrac{1}{4}}$(110) surface we have used a monoclinic $1\times2$ unit cell of $6.497 \times 5.916\times 40$ \AA$^3$, i.e.,
\begin{equation}
\left(\begin{array}{ccc}  \sqrt{2}a & c & 0\\ 0 & 2c & 0\\ 0 & 0 & \sqrt{2}{a} + D\end{array}\right),
\end{equation}
to maximize the separation between the   O$_{\textit{br}}^{\textit{vac}}$.  For the H$_2$O covered surfaces, we have employed a four layer slab with adsorbates on both sides and an orthorhombic $1\times 2$ unit cell of $6.497 \times 5.916 \times 47$~\AA$^3$, i.e.,
\begin{equation}
\left(\begin{array}{ccc}  \sqrt{2}a & 0 & 0\\ 0 & 2c & 0\\ 0 & 0 & \sqrt{2}{a} + D\end{array}\right),
\end{equation}
where $D\approx 34$~\AA.
We employed a $\Gamma$ centered $4\times4\times1$ \textbf{k}-point mesh, with approximately 9\sfrac{1}{6} unoccupied bands per atom, i.e.\ including all levels up to 30~eV above the VBM, an energy cutoff of 80 eV for the number of \textbf{G}-vectors, and a sampling of 80 frequency points for the dielectric function.   The $G_0W_0$ parameters are consistent with those previously  used  for describing rutile TiO$_2$ bulk, TiO$_2$(110) clean surface and interfaces\cite{OurJACS,MiganiLong}.  These parameters have been shown to provide accurate descriptions of bulk optical absorption spectra, and both clean surface and interfacial level alignment\cite{OurJACS,MiganiLong}.

\new{To model H$_2$O in the gas phase, we employed a unit cell with C$_{2v}$ symmetry and 16~\AA\ of vacuum in each direction.  At the $G_0W_0$ level, we used a smaller energy cutoff of 40~eV for the number of $\textbf{G}$-vectors, which has previously shown to provide an accurate description of the optical absorption spectra for isolated molecules \cite{VitoBenzeneBSE,VitoFullereneBSE}.}

To obtain DFT total energies and the relaxed structure of the clean reduced TiO$_{2-\sfrac{1}{8}}$(110) we have used a monoclinic $1\times4$ unit cell of $6.497 \times 11.832 \times 28$~\AA$^3$, i.e., 
\begin{equation}
  \left(\begin{array}{ccc}  \sqrt{2}a & 2c& 0\\  & 4c & 0\\ 0 & 0 & \sqrt{2}{a} + D\end{array}\right),
\end{equation}
where $D \approx 15$~\AA, and employed a $\Gamma$-centered $4\times 2 \times 1$ \textbf{k}-point mesh.

In this study, we have performed PBE and subsequent single-point RPBE\cite{RPBE} based DFT calculations for the H$_2$O adsorption energies $E_{\textit{ads}}$ on the stoichiometric and reduced surfaces.  The RPBE xc-functional was especially developed for the prediction of adsorption properties on metal surfaces \cite{RPBE}.   The H$_2$O adsorption energy on the Ti$_{\textit{cus}}$ site of a stoichiometric TiO$_2$(110) surface is given by
\begin{equation}
E_{\textit{ads}} \approx \frac{E[n\textrm{H}_2\textrm{O}+\textrm{TiO}_2\textrm{(110)}] - E[\textrm{TiO}_2\textrm{(110)}]}{n}- E[\textrm{H}_2\textrm{O}],
\end{equation}
where $n$ is the number of adsorbed H$_2$O functional units in the supercell, and $E[n\textrm{H}_2\textrm{O}+\textrm{TiO}_2\textrm{(110)}]$, $E[\textrm{TiO}_2\textrm{(110)}]$, and $E[\textrm{H}_2\textrm{O}]$ are the total energies of the covered and clean stoichiometric surfaces and gas phase water molecule, respectively.  Similarly, the H$_2$O adsorption energy on the O$_{\textit{br}}^{\textit{vac}}$ site of a reduced TiO$_{2-x}$(110) surface is given by
\begin{equation}
E_{\textit{ads}} \approx \frac{E[n\textrm{H}_2\textrm{O}+\textrm{TiO}_{2-x}\textrm{(110)}] - E[\textrm{TiO}_{2-x}\textrm{(110)}]}{n} - E[\textrm{H}_2\textrm{O}] ,
\end{equation}
where $E[n\textrm{H}_2\textrm{O}+\textrm{TiO}_{2-x}\textrm{(110)}]$ and $E[\textrm{TiO}_{2-x}\textrm{(110)}]$ are the total energies of the covered and clean reduced surfaces, respectively.

\newpage

\section{RESULTS AND DISCUSSION}

\subsection{Intact H$_{\text{2}}$O on the Stoichiometric Surface}\label{Sect:Intact}
\begin{figure}[!t]
%\noindent{\color{JCTCGreen}{\rule{\columnwidth}{1pt}}}
\includegraphics[width=\columnwidth]{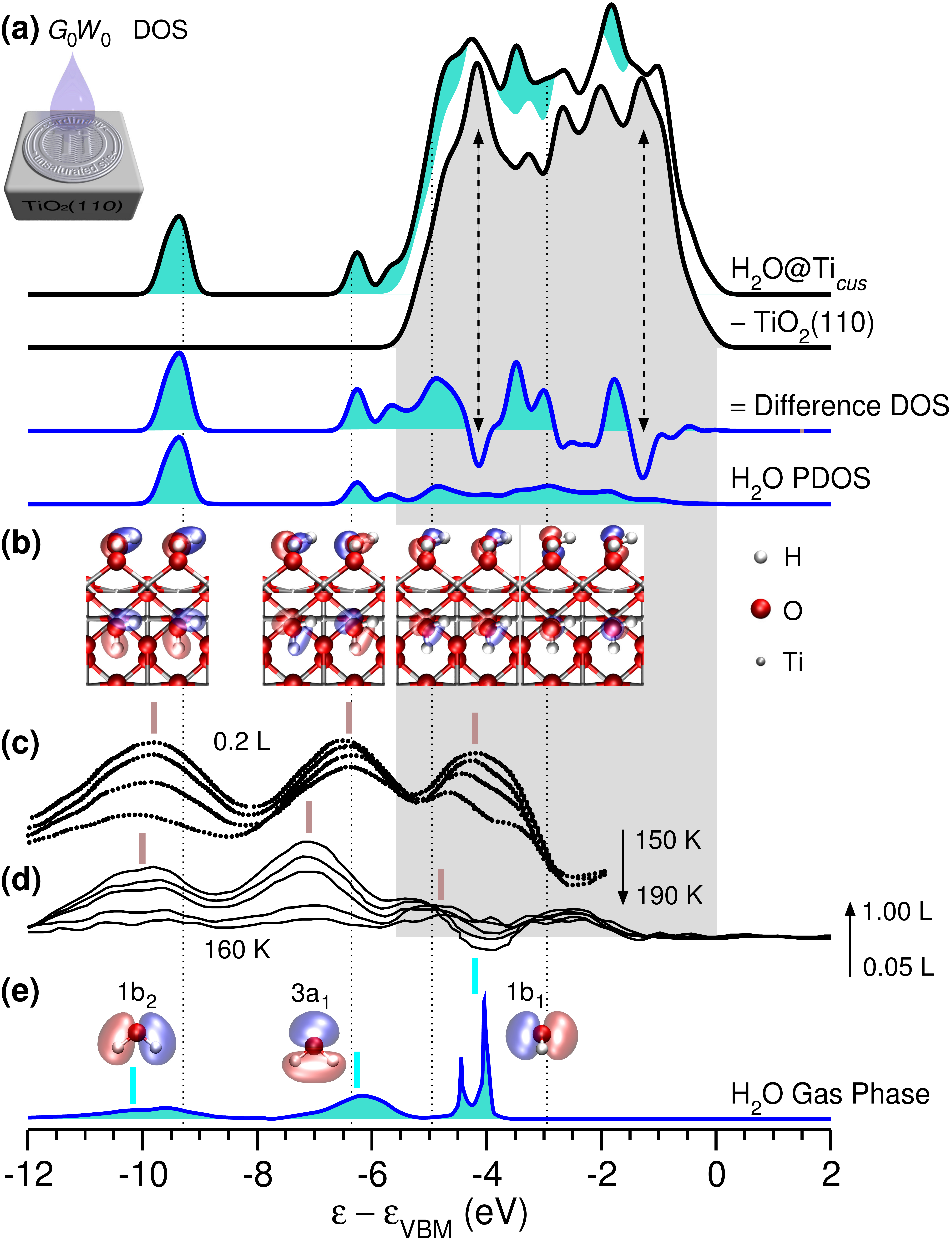}
\caption{Intact H$_{\text{2}}$O adsorbed \new{with parallel ($\rightrightarrows$) interfacial hydrogen bonds }on coordinately unsaturated Ti sites (H$_{\text{2}}$O@Ti$_{\textit{cus}}$).  \textbf{(a)} $G_0W_0$ DOS for 1 ML of intact H$_2$O covered (turquoise regions) or clean (gray region) stoichiometric TiO$_2$(110), their total DOS difference (dashed line), and the H$_2$O PDOS. \textbf{(b)} Selected molecular orbitals at $\Gamma$ and their energies (dotted lines). UPS difference spectra for H$_2$O covered TiO$_2$(110) \textbf{(c)} after 0.2 L exposure for $T = $ 150, 160, 175, and 190 K \cite{ThorntonH2ODissTiO2110} and \textbf{(d)} for $T = $ 160 K after 0.05, 0.1, 0.3, 0.7, and 1 L exposure \cite{DeSegovia}. Peak positions\cite{ThorntonH2ODissTiO2110,DeSegovia} are marked in brown. \textbf{(e)} H$_2$O molecular orbitals\new{, $G_0W_0$ calculated eigenenergies marked in cyan,} and \new{experimental }gas phase spectrum \new{aligned with the 1b$_1$ level of \textbf{(c)}}\cite{WaterSpectra}. Energies are relative to the VBM ($\varepsilon_{\textrm{VBM}}$). Intensity references are provided for $\varepsilon > \varepsilon_{\textrm{VBM}}$ when available.}\label{fgr:Fig1} 
\noindent{\color{JCTCGreen}{\rule{\columnwidth}{1pt}}}
\end{figure}

In Figure~\ref{fgr:Fig1} 
we disentangle adsorbate and substrate contributions to the spectrum of intact H$_{\text{2}}$O@Ti$_{\textit{cus}}$, and compare the H$_2$O PDOS to the theoretical and experimental difference DOS. Specifically, we model a monolayer (ML) of H$_2$O molecules \new{with parallel ($\rightrightarrows$) interfacial hydrogen bonds }aligned along the [001] direction (Figure~\ref{fgr:Fig1}\textbf{(b)})\cite{KennethJordanWaterChain,KimmelJPCL2012}.  Note that 1ML of intact H$_2$O is the most stable coverage and structure on the stoichiometric rutile TiO$_2$(110) surface \cite{MichaelidesDynamics}.    

The theoretical difference DOS is the difference between the total DOS of the H$_2$O covered (H$_{\text{2}}$O@Ti$_{\textit{cus}}$) and clean stoichiometric (TiO$_2$(110)) surfaces, as shown schematically in Figure~\ref{fgr:Fig1}\textbf{(a)}. Turquoise areas in the H$_2$O@Ti$_{\textit{cus}}$ and difference DOS indicate regions of greater density for the H$_2$O covered versus clean stoichiometric surface. The gray area indicates the DOS energy range for the clean stoichiometric TiO$_2$(110) surface. Figure~\ref{fgr:Fig1}\textbf{(c)} and \textbf{(d)} show two sets of UPS difference spectra obtained either by  raising the temperature (from 150 K to 190 K) for a consistent exposure to H$_2$O (0.2 L) for an  annealed TiO$_2$(110) surface\cite{ThorntonH2ODissTiO2110} (Figure~\ref{fgr:Fig1}\textbf{(c)}), or by  increasing the H$_2$O dose (from 0.01 L to 1 L) at low temperature (160 K) for a nearly perfect surface\cite{DeSegovia} (Figure~\ref{fgr:Fig1}d). The experimental spectra have been referenced to the VBM, which is positioned 3.2 eV below the experimental Fermi level.\cite{MiganiLong}  

Comparing the difference DOS to the H$_2$O PDOS, we find the peaks lying outside the TiO$_2$(110) DOS  energy range are clearly attributable to H$_2$O levels. As shown in Figure~\ref{fgr:Fig1}\textbf{(b)}, these levels are related to the 1b$_2$ and 3a$_1$ H$_2$O orbitals shown in Figure  \ref{fgr:Fig1}e. This is not the case within the TiO$_2$(110) DOS region, where the adsorbate levels are broadened by hybridization with the surface.  This hybridization with the surface has been severely underestimated by previous cluster-based MP2 calculations\cite{ClusterH2OTiO2CPL2003}. Within the TiO$_2$(110) DOS region, the peaks in the  H$_2$O PDOS have corresponding peaks in the difference DOS, although the relative peak intensities differ substantially between the two methods. More importantly, the difference DOS has dips centered at $-4.1$, $-2.4$, and $-1.1$~eV, where there are adsorbate levels in the PDOS, and a peak at $-0.4$~eV, where there are no adsorbate levels in the PDOS. The dips at $-4.1$ and $-1.1$~eV correspond to the O 2p$_\sigma$ and O 2p$_\pi$ peaks in the TiO$_2$(110) DOS \cite{DuncanTiO2}, respectively, as marked in Figure~\ref{fgr:Fig1}\textbf{(a)}. These peaks split due to mixing with the  3a$_1$ and 1b$_1$ H$_2$O orbitals. This splitting is the origin of the observed dips in the difference DOS, which are also seen experimentally in Figure~\ref{fgr:Fig1}\textbf{(c)} and d.

The peak at $-9.4$~eV in the H$_2$O PDOS, which has 1b$_2$ molecular character, agrees semi-quantitatively with the most strongly bound experimental peaks at $-9.8$~eV (Figure~\ref{fgr:Fig1}\textbf{(c)}) or $-10.0$~eV (Figure~\ref{fgr:Fig1}d). The peak at $-6.3$~eV in the H$_2$O PDOS, which has intermolecular 3a$_1$ bonding character, agrees semi-quantitatively with the experimental peaks at $-6.4$~eV (Figure~\ref{fgr:Fig1}\textbf{(c)}) or $-7.1$~eV (Figure~\ref{fgr:Fig1}d).  Note that the theoretical average deviation is  within that amongst the experiments. This may reflect differences in sample preparation, which result in a variety of different H$_2$O configurations, i.e., H$_2$O coverages, O$_{\textit{br}}^{\textit{vac}}$ concentrations, and mixtures of intact and dissociated H$_2$O. As we will show in Section~\ref{Sect:Dependence}, by considering a variety of H$_2$O structures a more complete description of the experiment is obtained. Altogether, this agreement for the $-9.4$ and $-6.3$~eV PDOS peaks lends confidence to our results for regions where the experimental results are unclear.   

The assignment of the peaks located within the TiO$_2$(110) DOS is much more complicated. The assumption that the highest peak in the experimental spectra originates solely from the H$_2$O 1b$_1$ level \cite{ThorntonH2ODissTiO2110,Krischok} is an oversimplification. In fact, both the 3a$_1$ and 1b$_1$ molecular levels contribute within this region (Figure~\ref{fgr:Fig1}\textbf{(b)}).   While the levels with intermolecular 3a$_1$ bonding character give rise to a distinct peak below the TiO$_2$(110) DOS region, those  with  intermolecular 3a$_1$ antibonding character are pushed to higher energies and mixed with the 1b$_1$ molecular levels (Figure~\ref{fgr:Fig1}\textbf{(b)}).  The latter is due to symmetry breaking at the interface.  Consequently, the H$_2$O PDOS is broadened into several peaks between $-5 $ and $-1$~eV.  These levels have interfacial (3a$_1$/1b$_1$-- O 2p$_{\sigma}$/2p$_\pi$) bonding and antibonding character (not visible at the isosurface value used).   

\subsection{Dissociated H$_{\text{2}}$O on Reduced Surfaces}\label{Sect:Dissociated}
\begin{figure}[!ht]
%{\color{JCTCGreen}{\rule{\columnwidth}{1.0pt}}}
\includegraphics[width=\columnwidth]{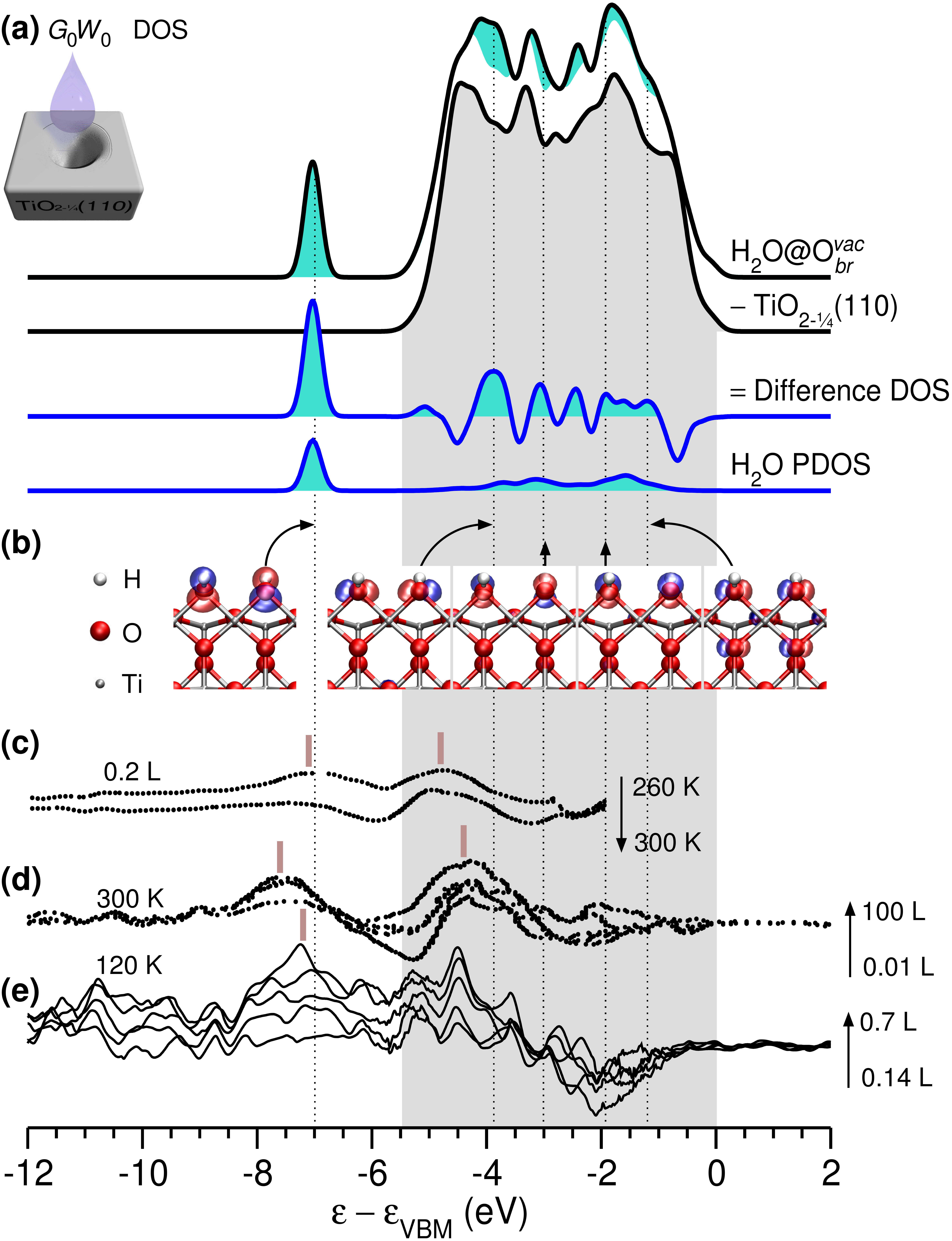}
\caption{H$_{\text{2}}$O dissociated on bridging O vacancies (H$_{\text{2}}$O@O$_{\textit{br}}^{\textit{vac}}$).  \textbf{(a)} $G_0W_0$ DOS for \sfrac{1}{2} ML of dissociated H$_2$O covered (turquoise regions) or clean (gray region) defective TiO$_{2-\sfrac{1}{4}}$(110) with \sfrac{1}{2}ML of O$_{\textit{br}}^{\textit{vac}}$, their total DOS difference (dashed line), and the H$_2$O PDOS. \textbf{(b)} Selected molecular orbitals and their energies (dotted lines). UPS difference spectra for H$_2$O on reduced TiO$_{2-x}$(110) \textbf{(c)} after 0.2 L exposure for $T = $ 260 and 300 K \cite{ThorntonH2ODissTiO2110},\textbf{(d)} for $T = $ 300 K after between 0.01 and 100 L exposure \cite{DeSegovia}, and \textbf{(e)} for $T = $ 120 K after 0.14, 0.3, 0.4, 0.5, and 0.7 L exposure \cite{Krischok}. Peak positions\cite{ThorntonH2ODissTiO2110,DeSegovia,Krischok} are marked in brown. Energies are relative to the VBM ($\varepsilon_{\textrm{VBM}}$). Intensity references are provided for $\varepsilon > \varepsilon_{\textrm{VBM}}$ when available.}\label{fgr:Fig2} 
{\color{JCTCGreen}{\rule{\columnwidth}{1.0pt}}}
\end{figure}

To see how dissociation of H$_2$O@O$_{\textit{br}}^{\textit{vac}}$ affects the spectrum, we now consider \sfrac{1}{2}ML of H$_2$O dissociated on a reduced TiO$_{2-\sfrac{1}{4}}$(110) surface (Figure~\ref{fgr:Fig2}). 
Here, we have used TiO$_{2-\sfrac{1}{4}}$(110) to denote a surface consisting of  \sfrac{1}{2}ML of O$_{\textit{br}}^{\textit{vac}}$ defects.  This structure corresponds to the staggered O$_{\textit{br}}$H surface species, shown in Figure~\ref{fgr:Fig2}\textbf{(b)}.

The theoretical difference DOS is the difference between the total DOS of the H$_2$O covered (H$_2$O@O$_{\textit{br}}^{\textit{vac}}$) and the clean reduced (TiO$_{2-\sfrac{1}{4}}$(110)) surfaces, shown schematically in Figure~\ref{fgr:Fig2}\textbf{(a)}. Turquoise areas in the
H$_2$O@O$_{\textit{br}}^{\textit{vac}}$ and difference DOS indicate regions of greater density
for the H$_2$O covered versus clean reduced surface. The gray area indicates
the DOS energy range for the clean reduced TiO$_{2-\sfrac{1}{4}}$(110) surface. The O$_{\textit{br}}^{\textit{vac}}$ defects give rise to occupied levels with Ti 3d character that are just below the conduction band minimum and outside the energy range shown.\cite{ThorntonRevTiO2110,JinZhaoJCP2009STM,DiValentinPRL} Note that  the H$_2$O PDOS includes half the O atoms and all the H atoms that make up the O$_{\textit{br}}$H species.  In this way the PDOS is provided in terms of H$_2$O formula units.

The peak in the difference DOS and PDOS at $-7.0$~eV has  O$_{\textit{br}}$H  $\sigma$ character, as shown in Figure~\ref{fgr:Fig2}\textbf{(b)}.  Note that the  peak intensity in the PDOS is  about half that in the difference DOS, as the PDOS includes half the O$_{\textit{br}}$ atoms. This peak's position agrees semi-quantitatively with the experimental peaks at $-7.1$ (Figure~\ref{fgr:Fig2}\textbf{(c)}), $-7.6$ (Figure~\ref{fgr:Fig2}d), or $-7.2$ eV (Figure~\ref{fgr:Fig2}e). The PDOS has a broader feature between $-4$ and $-1$ eV, due to hybridization with the surface. This feature is associated with contributions coming from the bonding and antibonding combinations of two distinct p orbitals of the O$_{\textit{br}}$H  species (Figure~\ref{fgr:Fig2}\textbf{(b)}): one  perpendicular to the   O$_{\textit{br}}$H  $\sigma$ bonds (the so-called OH $\pi$ level of NaOH \cite{NaOH}); the other  in the plane of the  O$_{\textit{br}}$H  $\sigma$ bonds. The lowest of these peaks at $-3.9$ eV corresponds to the bonding combination of the O$_{\textit{br}}$H  $\pi$ levels. 
This peak's position agrees semi-quantitatively with the consistently observed experimental peaks at $-4.8$,  $-4.4$, and $-4.5$~eV in Figures \ref{fgr:Fig2}\textbf{(c)}, \ref{fgr:Fig2}d, and \ref{fgr:Fig2}e, respectively.  However, the  antibonding O$_{\textit{br}}$H $\pi$ levels are shifted to much higher energies ($-1.2$~eV), as shown in Figure~\ref{fgr:Fig2}\textbf{(b)}. 

\begin{figure}%[!t]%[!htb]
%{\color{JCTCGreen}{\rule{\columnwidth}{1.0pt}}}
\includegraphics[width=\columnwidth]{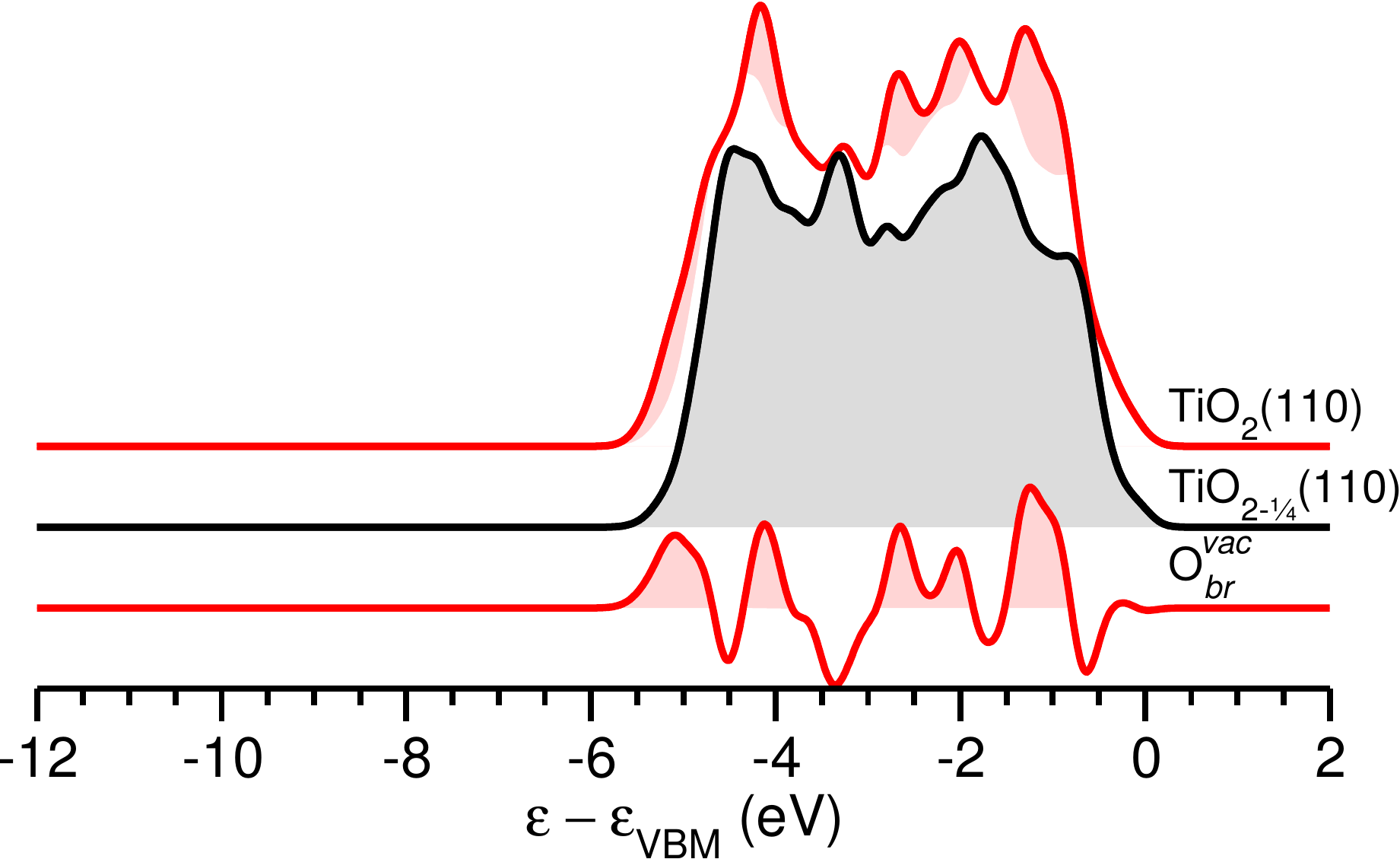}
\caption{O$_{\textit{br}}^{\textit{vac}}$ difference DOS between (red) stoichiometric TiO$_2$(110) and (black) reduced TiO$_{2-\sfrac{1}{4}}$(110) with \sfrac{1}{2}ML of O$_{\textit{br}}^{\textit{vac}}$ defects.  Red areas indicate defect healing of O$_{\textit{br}}^{\textit{vac}}$, i.e., regions of greater density for the stoichiometric versus reduced surfaces, shown in Figures~\ref{fgr:Structures_a} and \ref{fgr:Structures_c}, respectively.}\label{fgr:FigS2}
%{\color{JCTCGreen}{\rule{\columnwidth}{1.0pt}}}
\end{figure}

Much of the theoretical difference DOS's structure is attributable to the defect healing of O$_{\textit{br}}^{\textit{vac}}$, as seen from the difference DOS between TiO$_{2}$(110) and TiO$_{2-\sfrac{1}{4}}$(110) in Figure~\ref{fgr:FigS2}.  
This suggests that the observed features in the experimental difference spectra overlapping with the reduced surface's DOS are simply O$_{\textit{br}}$ levels reintroduced by dissociated H$_2$O@O$_{\textit{br}}^{\textit{vac}}$.  In particular, the peak which is usually attributed to O$_{\textit{br}}$H $\pi$ levels is actually composed of O$_{\textit{br}}$  surface levels unrelated to the presence of H atoms.

\subsection{\new{XC-Functional and Methodology Dependence of H$_{\text{2}}$O Spectra for Stoichiometric and Reduced Surfaces}}\label{Sect:xc-functionals}

\begin{figure}[!t]
%{\color{JCTCGreen}{\rule{\columnwidth}{1.0pt}}}
\includegraphics[width=\columnwidth]{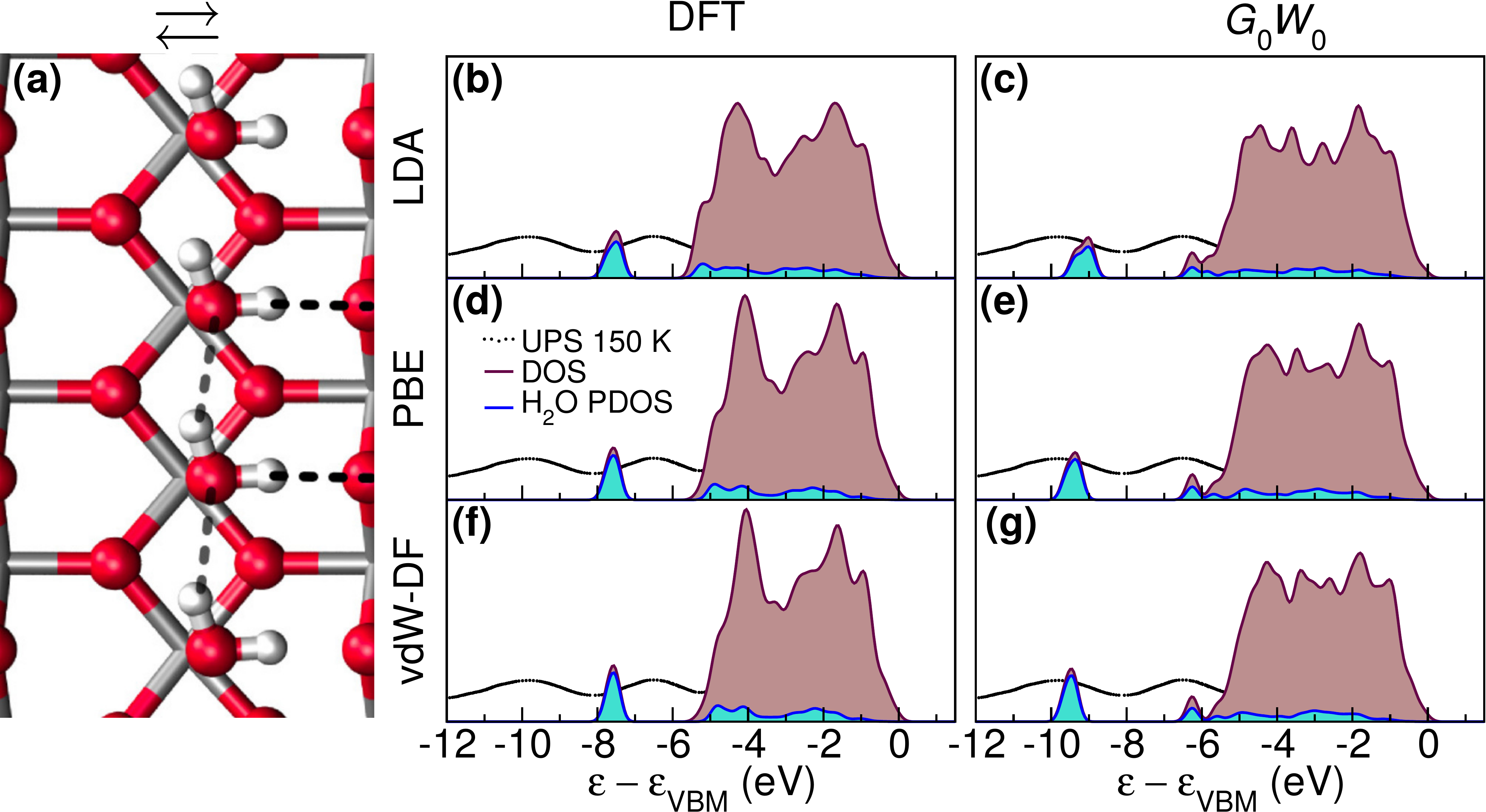}
\caption[H$_2$O@Ti$_{\textit{cus}}$]{\new{\textbf{(a)}} 1ML intact H$_{\text{2}}$O adsorbed \new{with parallel ($\rightrightarrows$) interfacial hydrogen bonds (black dashed lines) }on coordinately unsaturated Ti sites (H$_{\text{2}}$O@Ti$_{\textit{cus}}$). Total (maroon) and H$_2$O projected (blue) DOS  computed with \textbf{(\new{b,d,f})} DFT and \textbf{(\new{c,e,g})} $G_0W_0$ using the \textbf{(b\new{,c})} local density approximation (LDA)\cite{LDA} \textbf{(d\new{,e})} generalized gradient approximation (PBE)\cite{PBE} and \textbf{(f\new{,g})} long-ranged van der Waals interactions (vdW-DF)\cite{vdW-DF} for the xc-functional. The calculated H$_2$O PDOS are compared with the UPS spectrum at 150 K after 0.2 L exposure\cite{ThorntonH2ODissTiO2110}  (black).  Energies are relative to the valence band maximum, $\varepsilon_{\mathrm{VBM}}$.}\label{fgr:FigS1}
{\color{JCTCGreen}{\rule{\columnwidth}{1.0pt}}}
\end{figure}

\begin{figure}%[!t]
%{\color{JCTCGreen}{\rule{\columnwidth}{1.0pt}}}
\includegraphics[width=\columnwidth]{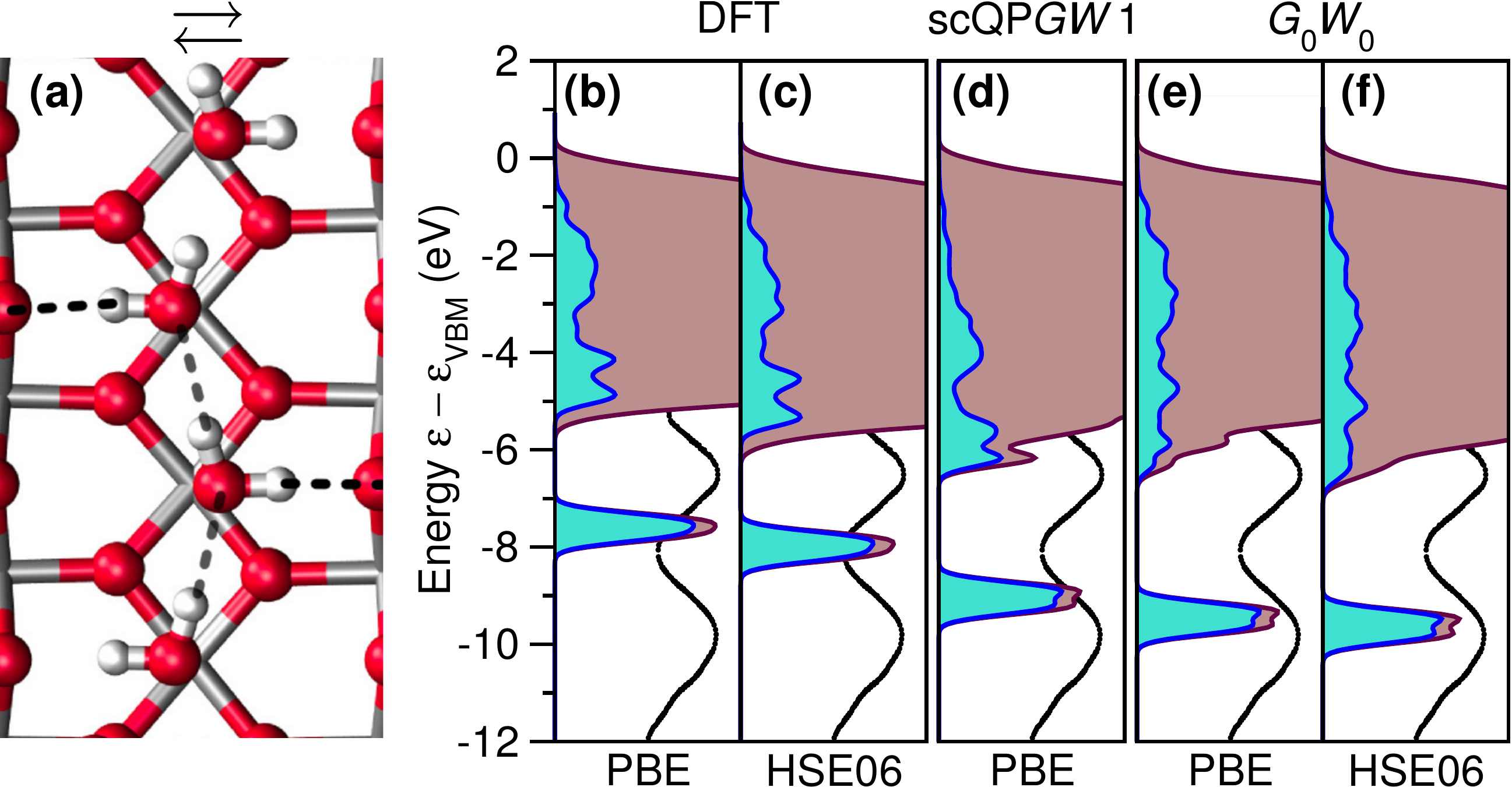}
\caption[H$_2$O@Ti$_{\textit{cus}}$]{
\new{\textbf{(a)} 1ML intact H$_{\text{2}}$O adsorbed with antiparallel ($\rightleftarrows$) interfacial hydrogen bonds on coordinately unsaturated Ti sites (H$_{\text{2}}$O@Ti$_{\textit{cus}}$). Total (maroon) and H$_2$O projected (blue) DOS  computed with \textbf{(b,c)} DFT, \textbf{(d)} scQP$GW$1, and \textbf{(e,f)} $G_0W_0$ using the \textbf{(b,d,e)} generalized gradient approximation (PBE)\cite{PBE} and \textbf{(c,f)} range-separated hybrid (HSE06)\cite{HSE06} for the xc-functional. The calculated H$_2$O PDOS are compared with the UPS spectrum at 150 K after 0.2 L exposure\cite{ThorntonH2ODissTiO2110}  (black).  Energies are relative to the valence band maximum, $\varepsilon_{\mathrm{VBM}}$.}
}\label{fgr:FigS1Antiparallel}
%{\color{JCTCGreen}{\rule{\columnwidth}{1.0pt}}}
\end{figure}

\begin{figure}[!t]
%{\color{JCTCGreen}{\rule{\columnwidth}{1.0pt}}}
\includegraphics[width=\columnwidth]{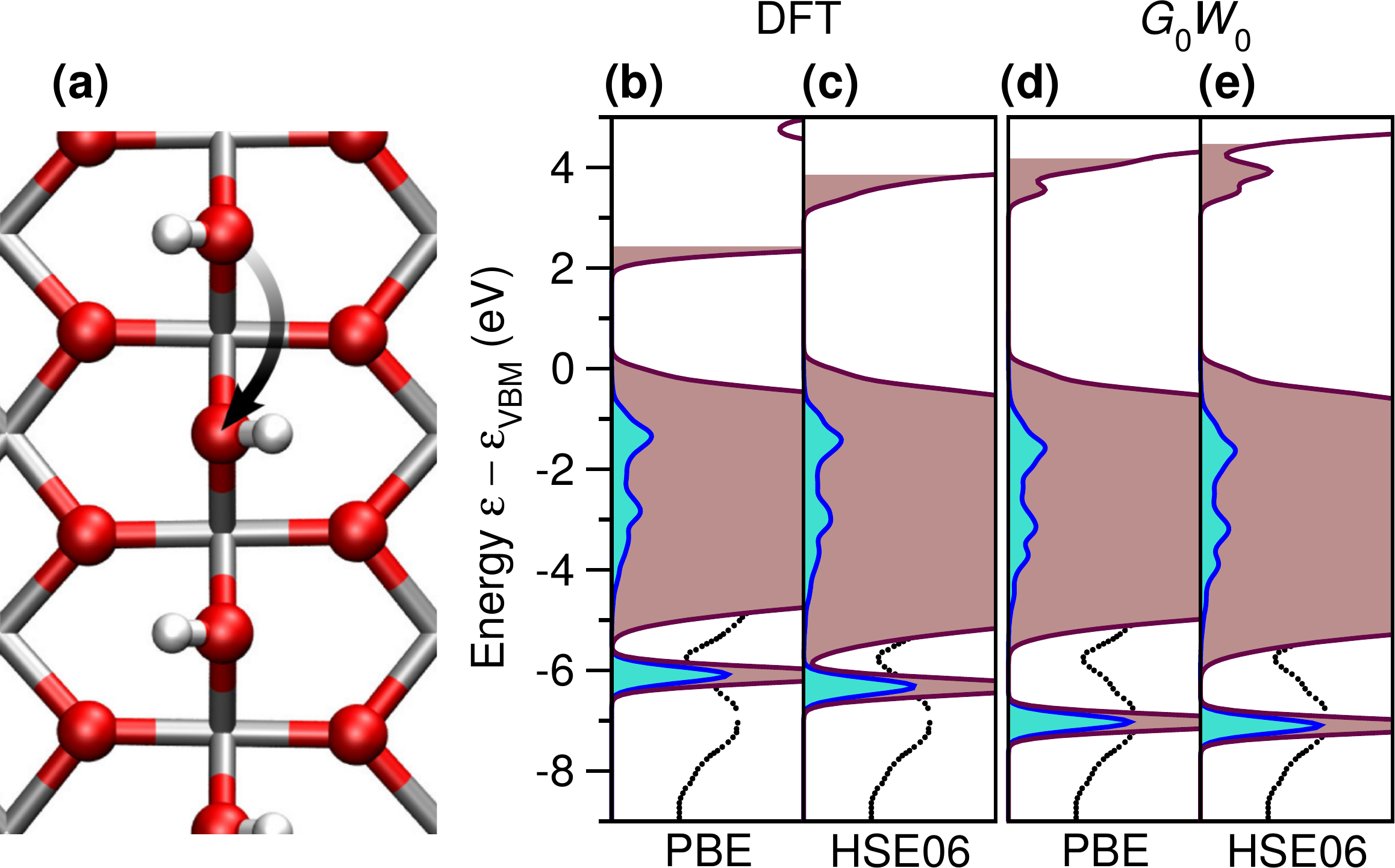}
\caption[H$_2$O@Ti$_{\textit{cus}}$]{
\new{\textbf{(a)} \sfrac{1}{2}ML H$_{\text{2}}$O dissociated on bridging O vacancies (H$_{\text{2}}$O@O$_{\textit{br}}^{\textit{vac}}$) of defective TiO$_{2-\text{\sfrac{1}{4}}}$(110) with \sfrac{1}{2}ML of O$_{\textit{br}}^{\textit{vac}}$. Total (maroon) and H$_2$O projected (blue) DOS  computed with \textbf{(b,c)} DFT and \textbf{(d,e)} $G_0W_0$ using the \textbf{(b,d)} generalized gradient approximation (PBE)\cite{PBE} and \textbf{(c,e)} range-separated hybrid (HSE06)\cite{HSE06} for the xc-functional. Filling denotes occupation. The calculated H$_2$O PDOS are compared with the UPS spectrum at 300 K after 0.2 L exposure\cite{ThorntonH2ODissTiO2110}  (black).  Energies are relative to the valence band maximum, $\varepsilon_{\mathrm{VBM}}$.}
}\label{fgr:FigReducedHSE}
{\color{JCTCGreen}{\rule{\columnwidth}{1.0pt}}}
\end{figure}

\old{In Figures~\ref{fgr:FigS1} and \ref{fgr:S1Antiparallel}} 
\new{To assess the robustness of the calculated QP H$_2$O PDOS, we consider its dependence on the xc-functional and methodology.  Specifically, }we compare the H$_2$O PDOS \old{computed with}\new{from} \old{\textbf{(a,c,e)} }DFT\new{, scQP$GW$1,} and \old{\textbf{(b,d,f)} }$G_0W_0$ for 1ML intact H$_2$O@Ti$_{\textit{cus}}$\new{ with parallel ($\rightrightarrows$) and antiparallel ($\rightleftarrows$) interfacial hydrogen bonds and \sfrac{1}{2}ML dissociated H$_2$O@O$_{\textit{br}}^{\textit{vac}}$ in Figures~\ref{fgr:FigS1}, \ref{fgr:FigS1Antiparallel}, and \ref{fgr:FigReducedHSE}, respectively}.
%{\color{JCTCGreen}{\rule{\columnwidth}{1.0pt}}}
\begin{table}[!t]
%\noindent{\color{JCTCGreen}{\rule{\columnwidth}{1.0pt}}}
\caption[Height of 1ML I H$_{\text{2}}$O@Ti$_{\textit{cus}}$]{\textbf{\textrm{Height of H$_{\text{2}}$O Above TiO$_{\text{2}}$(110) for 1ML Intact H$_{\text{2}}$O@Ti$_{\textit{cus}}$ Measured with SXPS and Calculated with LDA, PBE, or vdW-DF XC-Functionals.}}
}\label{table:TS1}
\begin{tabular}{lc}
\multicolumn{2}{>{\columncolor[gray]{.85}}c}{ }\\[-3mm]
\rowcolor[gray]{.85}Method & \multicolumn{1}{>{\columncolor[gray]{.85}}c}{$d[\textrm{H}_2\textrm{O}-\textrm{Ti}_{\textit{cus}}]$ (\AA)} \\[1mm]
SXPS\cite{H2O-TiO2110_distancePRL,H2O-TiO2110_distanceSurfSci} & 2.210\\
LDA & 2.180\\
PBE & 2.367\\
vdW-DF & 2.434 
\end{tabular}
\noindent{\color{JCTCGreen}{\rule{\columnwidth}{1.0pt}}}
\end{table}

 We find the observed structure of the $G_0W_0$ H$_2$O PDOS is independent of whether the local density approximation (LDA)\cite{LDA}, generalized gradient approximation (PBE)\cite{PBE}, \old{or }long-ranged van der Waals interactions (vdW-DF)\cite{vdW-DF}\new{, or a range-separated hybrid (HSE06)\cite{HSE06}} are employed for the xc-functional.   \new{This is consistent with the previously reported similarities between PBE and HSE based $G_0W_0$ PDOS for CH$_3$OH on TiO$_2$(110)\cite{MiganiLong}.}  This is despite the greater differences observed amongst the  DFT H$_2$O PDOS, which all differ qualitatively from the experiments.  
Furthermore,  the $G_0W_0$ H$_2$O PDOS is robust to the resulting changes in the H$_2$O height above the surface, i.e., the distance between H$_2$O and Ti$_{\textit{cus}}$ $d[\textrm{H}_2\textrm{O}-\textrm{Ti}_{\textit{cus}}]$, shown in Table~\ref{table:TS1}.
\new{Furthermore, Figure~\ref{fgr:FigS1Antiparallel}\textbf{(d,e)} shows that scQP$GW$1 provides a similar H$_2$O PDOS level alignement to $G_0W_0$.   This is consistent with what was previously reported for the CH$_3$OH--TiO$_2$(110) interface \cite{OurJACS,MiganiLong}.
}

We clearly see that the differences between the DFT and $G_0W_0$ PDOS, i.e., the QP energy shifts, are far from simply being rigid.    For instance, we find for \old{\textbf{(c,d)}} PBE that the QP energy shifts for the levels that contribute to the highest-energy PDOS peak $\varepsilon_{\textit{peak}}^{\textit{PDOS}}$ are almost negligible\new{ (\emph{cf.} Figures \ref{fgr:FigS1}\textbf{(d,e)} and \ref{fgr:FigS1Antiparallel}\textbf{(b,e)})} .  As a result,  the QP $G_0W_0$ $\varepsilon_{\textit{peak}}^{\textit{PDOS}}$ is only  $\sim0.1$~eV lower compared to DFT.  On the other hand, we find significant QP shifts to stronger binding  for the levels that contribute to the most strongly bound PDOS peak  with 1b$_2$ $\sigma$ molecular character. \old{As a result}\new{For example}, with PBE the QP $G_0W_0$ \new{lowest energy }peak\old{ at $-9.4$~eV} is shifted by $\sim-1.7$~eV compared to DFT\new{ (\emph{cf.} Figures \ref{fgr:FigS1}\textbf{(d,e)} and \ref{fgr:FigS1Antiparallel}\textbf{(b,e)})}.

 As previously shown for the CH$_3$OH--TiO$_2$(110) interface,  these differences in the shifts of the peaks  are directly related to differences in the spatial distribution of the wave functions for the levels contributing to the  peaks.\cite{OurJACS,MiganiLong,MiganiInvited} This is because the QP $G_0W_0$ corrections to the DFT eigenenergies for interfaces are directly correlated with the spacial distribution of the wave functions.\cite{OurJACS,MiganiLong,MiganiInvited} The 
negligible shift of the \old{PBE}\new{DFT} highest-energy PDOS peak\new{ (Figures~\ref{fgr:FigS1} \textbf{(b,d,f)} and \ref{fgr:FigS1Antiparallel}\textbf{(b,c)})}  is due to its strong hybridization with the surface, i.e., weight on TiO$_2$(110), for the levels contributing to this peak.\cite{OurJACS,MiganiLong,MiganiInvited}  On the other hand, the levels that contribute to the most strongly bound PDOS peak have little weight on TiO$_2$(110), and have $\sigma$ character.  Both their localized H$_2$O character as well as their $\sigma$  nature explain why these levels have large QP energy shifts to stronger binding.\cite{OurJACS,MiganiLong,MiganiInvited}

\new{Oxygen defective and hydroxylated ($h-$)TiO$_2$ surfaces have occupied 3d levels which are associated with reduced Ti$^{3+}$ atoms \cite{DiValentinPRL}. 
One such example is the \sfrac{1}{2}ML dissociated H$_2$O@O$_{\textit{br}}^{\textit{vac}}$ on reduced TiO$_{2-\text{\sfrac{1}{4}}}$(110) with \sfrac{1}{2}ML of O$_{\textit{br}}^{\textit{vac}}$ shown in Figure~\ref{fgr:FigReducedHSE}\textbf{(a)}.}
\new{The spacial distribution of the 3d density for O defective surfaces has been characterized by low temperature scanning tunneling microscopy (STM)\cite{JinZhaoJCP2009STM,Papageorgiou09022010}. STM measurements find at 77 K the 3d density is homogeneously distributed along the [001] direction,\cite{JinZhaoJCP2009STM}  while at $\sim5$~K the 3d density exhibits an asymmetric localized character.\cite{Papageorgiou09022010}}

\new{A localized description of the Ti$^{3+}$ occupied 3d levels is not obtained from DFT with standard xc-functionals.  For example, the occupied 3d levels obtained with PBE are highly delocalized, as clearly shown in Figure~\ref{fgr:FigReducedHSE}\textbf{(b)}. This is due to self-interaction errors which are inherent in such xc-functionals. If one performs spin-polarized DFT calculations with a hybrid xc-functional on such systems, one obtains localized Ti$^{3+}$ 3d$^1$ levels between 0.\old{9}\new{7} and 1.6~eV below the CBM, along with a structural deformation of the TiO$_2$(110) surface\cite{DiValentinPRL,JinZhaoJCP2009STM}.  However, spin-paired calculations with HSE06 on the PBE relaxed geometry only yield an occupied shoulder at the CBM (Figure~\ref{fgr:FigReducedHSE}\textbf{(c)}).  At the QP $G_0W_0$ level based on PBE, this shoulder evolves into a distinct peak about 0.6~eV below the Fermi level, $\varepsilon_{\textrm{F}}$.  This effect is even more pronouced when the $G_0W_0$ calculation is based on HSE06 (\emph{cf}. Figure~\ref{fgr:FigReducedHSE}\textbf{(d,e)}), which yields peaks at 0.6 and 0.9~eV below $\varepsilon_{\textrm{F}}$.}
\new{As compared to $G_0W_0$ PBE, $G_0W_0$ HSE06 shifts the unoccupied 3d levels further up in energy revealing the double peak structure. These energies are in very good agreement with the peak at 0.8~eV below $\varepsilon_{\textrm{F}}$ in the UPS spectra of H$_{\text{2}}$O@O$_{\textit{br}}^{\textit{vac}}$ of Figure~\ref{fgr:Fig2}\textbf{(d)}. This peak is not shown in Figure~\ref{fgr:Fig2}\textbf{(d)} as it is  slightly above 2 eV with respect to VBM.\cite{DeSegovia} However, note that $G_0W_0$ overestimates by about 1~eV the VBM position relative to $\varepsilon_{\textrm{F}}$ as compared with UPS experiments.\cite{DeSegovia}}

 \new{This}\new{ result}\new{ is completely independent of the wavefunction's spacial distribution, i.e., localization, as the $G_0W_0$ calculations are based on the KS wavefunctions.} \new{This is different from previous findings, which showed DFT with either PBE or hybrid xc-functionals is only giving  distinct peaks for the occupied 3d levels provided the relaxed spin-polarized distorted structure is used in the calculations.\cite{DiValentinPRL,JinZhaoJCP2009STM}}

\new{While for $G_0W_0$ based on PBE and HSE06 one sees noticeable differences in the description of the 3d occupied levels, the QP H$_2$O PDOS and its alignment relative to the VBM are unchanged.  Although localization of the Ti$^{3+}$ occupied levels and associated structural deformations are absent from our approach, such features should not significantly alter the QP H$_2$O PDOS.  This is because the Ti$^{3+}$ levels are too far above the VBM ($\sim 2$~eV\cite{DiValentinPRL}) to hybridize with the H$_2$O.  Moreover, as we will show in Section~\ref{Sect:Dependence}, the QP H$_2$O PDOS is rather robust to local deformations of the surface structure, e.g., due to changes in coverage.}

\old{ Finally, oxygen defective and hydroxylated surface structures obtained from DFT with a generalized gradient approximation xc-functional  give simulated scanning tunneling microscopy (STM) images, which agree with the STM measurements. This is in contrast to the simulated STM images based on the distorted structures with localized 3d electrons, which do not agree with the STM measurements.\cite{JinZhaoJCP2009STM}}

\subsection{Coverage and Dissociation Dependence of H$_{\text{2}}$O Spectra \old{on}\new{for} Stoichiometric and Reduced Surfaces}\label{Sect:Dependence}

\begin{figure}[!thb]
%\noindent{\color{JCTCGreen}{\rule{\columnwidth}{1pt}}}
\includegraphics[width=\columnwidth]{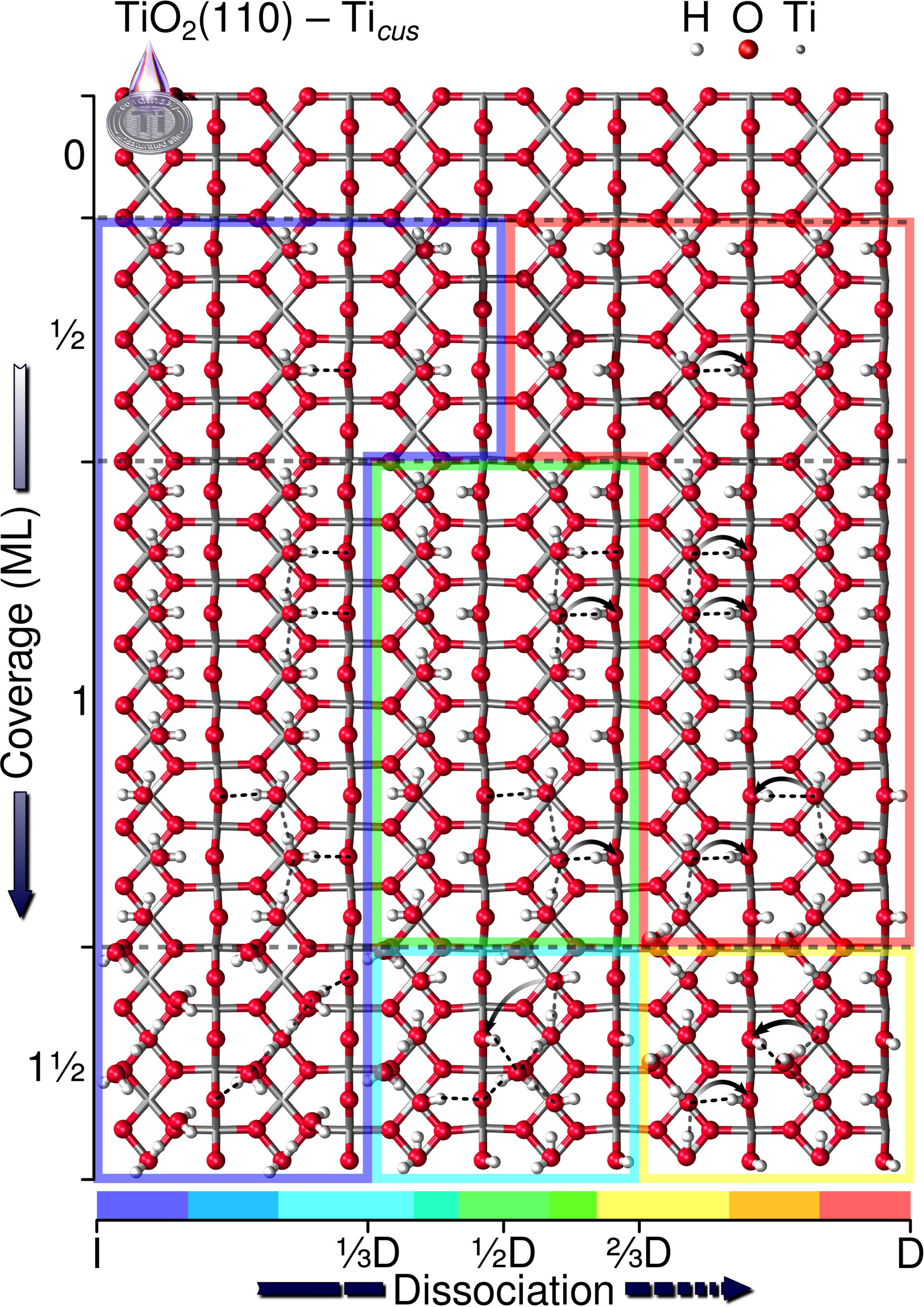}
\caption{{Schematics of H$_{\text{2}}$O adsorbed intact (I) or dissociated (D) on  coordinately unsaturated Ti sites (Ti$_{\textit{cus}}$) of stoichiometric TiO$_{\text{2}}$(110).} Higher coverages are obtained by the addition of second-layer H$_2$O.  Coverage is the number of H$_2$O formula units per (110) $1\times1$ unit area of the clean stoichiometric surface. Dissociation is the fraction of H$_2$O molecules which are dissociated, i.e., one minus the ratio of intact H$_2$O molecules to H$_2$O formula units. Colored frames encompass regions of common fractional dissociation. Charge transfer of about $-0.4e$ accompanying deprotonation\cite{OurJACS} of intact H$_2$O adsorbed at Ti$_{\textit{cus}}$ is represented by arrows, while intermolecular (gray) and interfacial (black) hydrogen bonds are denoted by dotted lines.}\label{fgr:Structures_a} 
\noindent{\color{JCTCGreen}{\rule{\columnwidth}{1pt}}}
\end{figure}

\begin{figure*}[!t]
\includegraphics[width=0.75\textwidth]{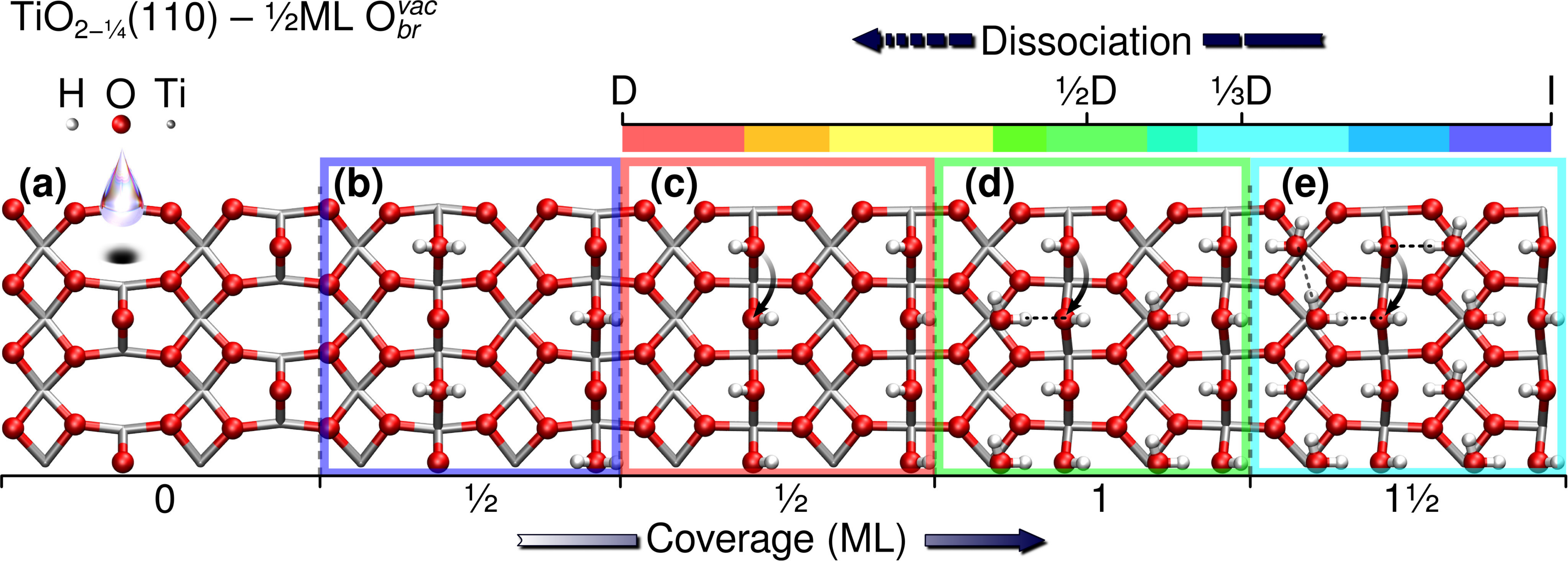}
\caption{Schematics of reduced TiO$_{\text{2}-\sfrac{1}{4}}$(110) with \sfrac{1}{2}ML of bridging O vacancies (O$_{\textit{br}}^{\textit{vac}}$)  \textbf{(a)} clean, covered with \sfrac{1}{2}ML \textbf{(b)} intact and \textbf{(c)} dissociated H$_{\text{2}}$O@O$_{\textit{br}}^{\textit{vac}}$, and with an additional \textbf{(d)} \sfrac{1}{2}ML or \textbf{(e)} 1ML of intact H$_{\text{2}}$O adsorbed on coordinately unsaturated Ti sites (Ti$_{\textit{cus}}$). Coverage is the number of H$_2$O formula units per (110) $1\times1$ unit area of the clean reduced surface. Dissociation is the fraction of H$_2$O molecules which are dissociated, i.e., one minus the ratio of intact H$_2$O molecules to H$_2$O formula units.  Charge transfer of about $-0.4e$ accompanying deprotonation\cite{OurJACS} is represented by arrows, while intermolecular (gray) and interfacial (black) hydrogen bonds are denoted by dotted lines.}\label{fgr:Structures_c} 
\noindent{\color{JCTCGreen}{\rule{\textwidth}{1pt}}}
\end{figure*}

As different experimental conditions and surface preparations have been employed, there are expected to be different H$_2$O structures on the surface.  To evaluate how strongly the DOS depends on the adsorption geometry, we now consider a variety of coverages of intact 
and dissociated H$_2$O on rutile stoichiometric TiO$_2$(110) 
(Figure~\ref{fgr:Structures_a})  and reduced TiO$_{2-\sfrac{1}{4}}$(110) (Figure~\ref{fgr:Structures_c})
\begin{figure}[!t]
%\noindent{\color{JCTCGreen}{\rule{\columnwidth}{1pt}}}
\includegraphics[width=0.765\columnwidth]{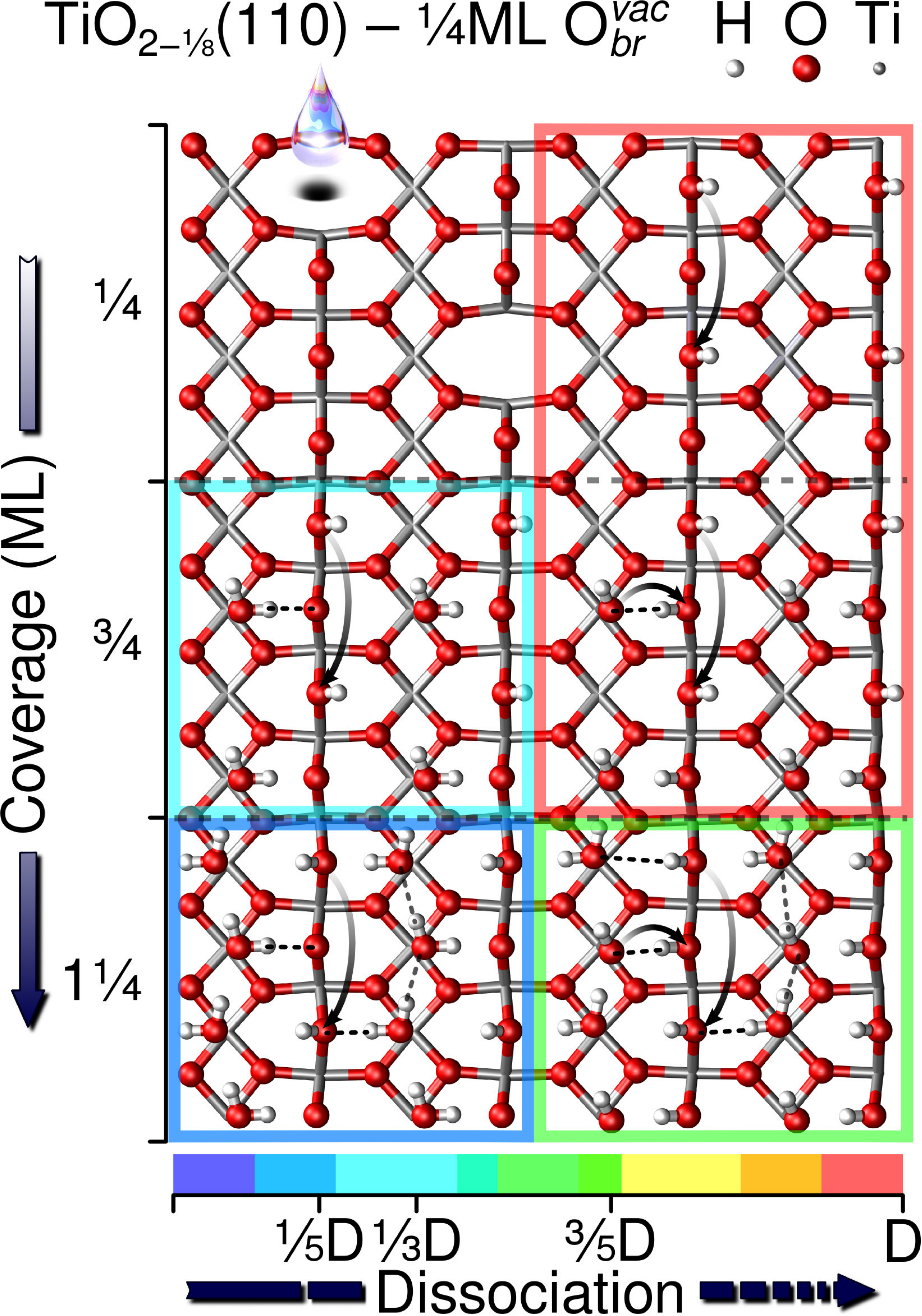}
\caption{{Schematics of H$_{\text{2}}$O adsorbed dissociated (D) on  \sfrac{1}{4}ML of bridging O vacancies (O$_{\textit{br}}^{\textit{vac}}$) on reduced TiO$_{\text{2}-x}$(110) ($x=$~\sfrac{1}{8}).} Higher coverages are obtained by the addition of H$_2$O@Ti$_{\textit{cus}}$.  Coverage is the number of H$_2$O formula units per (110) $1\times1$ unit area of the clean stoichiometric or reduced surface. Dissociation is the fraction of H$_2$O molecules which are dissociated, i.e., one minus the ratio of intact H$_2$O molecules to H$_2$O formula units. Colored frames encompass regions of common fractional dissociation. Charge transfer of about $-0.4e$ accompanying deprotonation\cite{OurJACS} of intact H$_2$O adsorbed at Ti$_{\textit{cus}}$ or O$_{\textit{br}}^{\textit{vac}}$ is represented by arrows, while intermolecular (gray) and interfacial (black) hydrogen bonds are denoted by dotted lines.}\label{fgr:Structures_b} 
\noindent{\color{JCTCGreen}{\rule{\columnwidth}{1pt}}}
\end{figure}
 and TiO$_{2-\sfrac{1}{8}}$(110) (Figure~\ref{fgr:Structures_b}) with \sfrac{1}{2}ML and \sfrac{1}{4}ML of O$_{\textit{br}}^{\textit{vac}}$ defects, respectively. 
  The relative importance of these geometries is illustrated in Figure~\ref{fgr:Spectra}\textbf{(a)} and \ref{fgr:Spectra}\textbf{(b)} by the average absorption energy $E_{\textit{ads}}$ per H$_2$O molecule on the stoichiometric or reduced surfaces\cite{SelloniWaterReview2010} with either PBE\cite{PBE} or RPBE\cite{RPBE} xc-functionals. In so doing, the contribution of different structures to the measured spectra can be disentangled.    Note that an intact \sfrac{1}{2}ML of H$_2$O@O$_{\textit{br}}^{\textit{vac}}$ (Figure~\ref{fgr:Structures_c}\textbf{(b)}) is probably only a transient locally stable state of the reduced H$_2$O--TiO$_{2-\text{\sfrac{1}{4}}}$(110) interface\cite{NorskovVacanciesPRL2001}, which may easily evolve into the $\sim0.7$~eV more stable dissociated \sfrac{1}{2}ML H$_2$O@O$_{\textit{br}}^{\textit{vac}}$ (Figure~\ref{fgr:Structures_c}\textbf{(c)}).  For this reason, we only consider dissociated H$_2$O@O$_{\textit{br}}^{\textit{vac}}$ structures in Figure~\ref{fgr:Spectra}d.

By comparing to lower coverage H$_2$O structures (\sfrac{1}{2}ML\cite{WaterTiO2ControversyPRL2004,LindanMonolayer,JinH2O,MichaelidesDynamics} to 1ML\cite{WaterTiO2ControversyPRL2004,LindanMonolayer,JinH2O,MichaelidesDynamics}  in Figure~\ref{fgr:Structures_a} and \sfrac{1}{4}ML\cite{0.25DObrvac} in Figure~\ref{fgr:Structures_b} to \sfrac{1}{2}ML\cite{WaterReducedTiO2110PRB2009} in Figure~\ref{fgr:Structures_c}), we can disentangle the effect of interaction between the H$_2$O molecules on the spectra.  Further, these structures allow us to probe the isolated molecule limit.  

As shown in Figure~\ref{fgr:Spectra}, at lower coverages  the overall width of the spectra is reduced with fewer distinct peaks.  When the coverage is increased to include intermolecular interactions between adjacent species, the molecular levels hybridize into bonding and antibonding intermolecular levels.  This produces  additional peaks above and below those present at low coverage.  As a result, the peak with intermolecular bonding 3a$_1$ character at $-6.3$~eV for 1ML of H$_2$O@Ti$_{\textit{cus}}$ is absent for a \sfrac{1}{2}ML coverage.  This reinforces the assignment of the experimental spectra shown in Figure~\ref{fgr:Fig1} to an intact 1ML H$_2$O@Ti$_{\textit{cus}}$ geometry with interacting molecules.

To see how the spectra for dissociation of H$_2$O@Ti$_{\textit{cus}}$ compare to H$_{2}$O@O$_{\textit{br}}^{\textit{vac}}$, we have considered the half-dissociated (\sfrac{1}{2}D) and fully dissociated (D) H$_2$O structures shown in Figure~\ref{fgr:Structures_a}.  As shown in Figure~\ref{fgr:Spectra}\textbf{(c)}, the peak at $-7.0$~eV with O$_{\textit{br}}$H $\sigma$ character for H$_{2}$O@O$_{\textit{br}}^{\textit{vac}}$ splits into two peaks for dissociated H$_2$O@Ti$_{\textit{cus}}$.  The lower energy peak has both O$_{\textit{cus}}$H and O$_{\textit{br}}$H $\sigma$ character, while the higher energy peak is mostly O$_{\textit{cus}}$H in character.  Furthermore, we find a similar couple of peaks for \sfrac{3}{4}ML mixtures of dissociated H$_2$O@Ti$_{\textit{cus}}$ and H$_{2}$O@O$_{\textit{br}}^{\textit{vac}}$ shown in Figure~\ref{fgr:Spectra}d.  This means one may recognize dissociated H$_2$O@Ti$_{\textit{cus}}$  by both the presence of two peaks at about $-7.0$ and $-6.3$~eV, and the absence of the low-energy peak with 1b$_2$ character for intact H$_2$O@Ti$_{\textit{cus}}$.

The absence of a peak at about $-6.3$~eV  in the experimental spectra shown in Figure~\ref{fgr:Fig2}\textbf{(c)} reinforces its attribution to dissociated H$_2$O@O$_{\textit{br}}^{\textit{vac}}$ rather than dissociated  H$_2$O@Ti$_{\textit{cus}}$.  This is further supported by the calculated H$_2$O absorption energies (Figure~\ref{fgr:Spectra}\textbf{(a)} and \ref{fgr:Spectra}\textbf{(b)}). These are generally weaker for dissociated H$_2$O@Ti$_{\textit{cus}}$, and stronger for H$_2$O@O$_{\textit{br}}^{\textit{vac}}$, as in previous calculations \cite{NorskovVacanciesPRL2001}.

To check whether changes in the absorption geometry of H$_2$O affect the spectra for the same coverage, we compare 1ML of H$_2$O \{I, \sfrac{1}{2}D, D\} adsorbed with either parallel \new{($\rightrightarrows$) }or antiparallel \new{($\rightleftarrows$) }interfacial hydrogen bonds \cite{KennethJordanWaterChain} (black dashed lines in Figure~\ref{fgr:Structures_a}).  Overall, the two sets of spectra are consistent, and demonstrate the general robustness of the DOS to minor changes in the water absorption geometry.  However, as the H$_2$O molecules are no longer equivalent when the interfacial hydrogen bonds are antiparallel, there is a greater splitting between bonding and antibonding contributions for the peaks with 1b$_2$ and 3a$_1$ molecular character.  In particular, for intact H$_2$O, the lowest energy peak with molecular 1b$_2$ character splits with a separate peak at $-9.6$~eV, which is closer to the peaks at $-9.8$\cite{ThorntonH2ODissTiO2110} (Figure~\ref{fgr:Fig1}\textbf{(c)}) and $-10.0$~eV\cite{DeSegovia} (Figure~\ref{fgr:Fig1}d) observed experimentally.

To see how increasing the H$_2$O coverage impacts the spectra, we compare monolayer (\sfrac{1}{2}ML or 1ML) to multilayer (1\sfrac{1}{2}ML) H$_2$O \{I, \sfrac{1}{3}D, \sfrac{2}{3}D\}\cite{LindanMultilayer} (Figure~\ref{fgr:Structures_a}), and consider the effect of additional H$_2$O@Ti$_{\textit{cus}}$ to \sfrac{1}{4}ML (Figure~\ref{fgr:Structures_b}) and \sfrac{1}{2}ML (Figure~\ref{fgr:Structures_c}) H$_2$O@O$_{\textit{br}}^{\textit{vac}}$\cite{JinH2O}.   In this way we can can see how robust the observed features in the individual spectra for isolated species are to screening by H$_2$O layers\cite{NakamuraJACS2004,ImanishiJACS2007}, and probe the liquid water limit\cite{LiquidWaterGW}.

\begin{figure*}[!t]
\includegraphics[width=\textwidth]{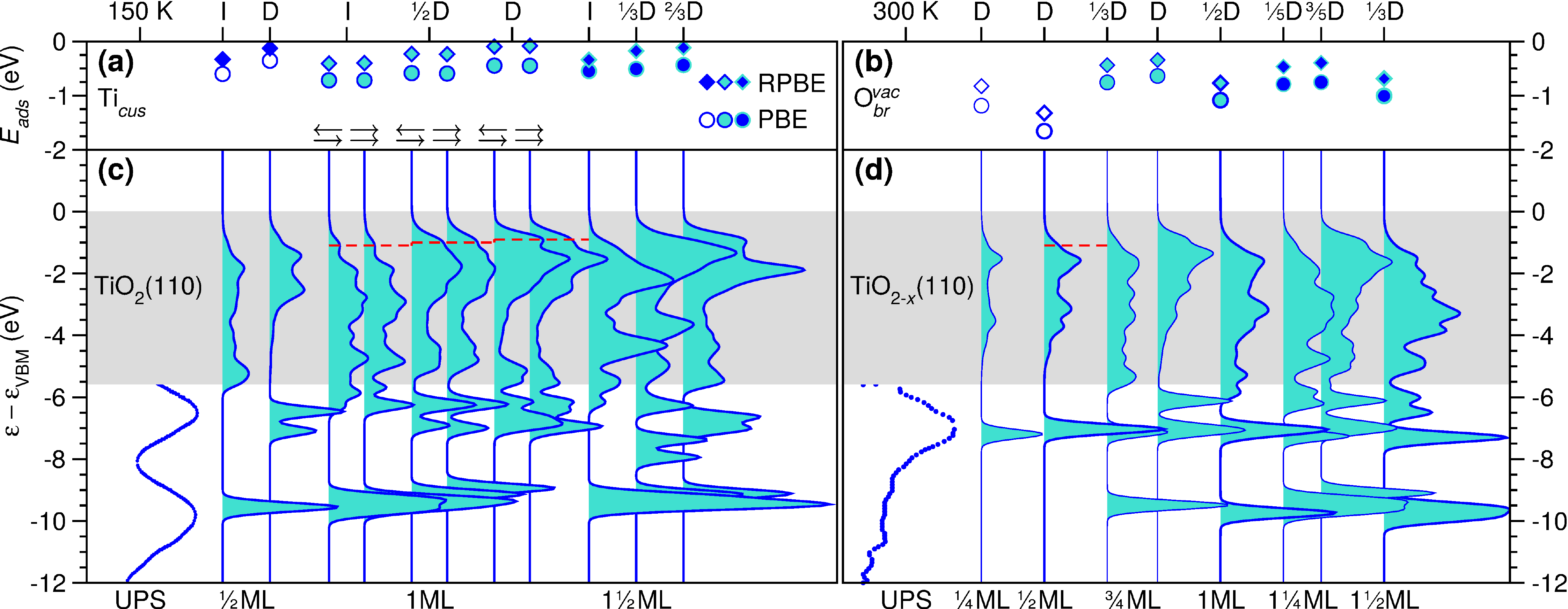}
\caption{
Structure and coverage dependence of \textbf{(a,b)} adsorption energy $E_{\textit{ads}}$ and \textbf{(c,d)} $G_0W_0$ PDOS for H$_{\text{2}}$O adsorbed intact (I) or dissociated (D) on  \textbf{(a,c)} coordinately unsaturated Ti sites (Ti$_{\textit{cus}}$) of stoichiometric TiO$_{\text{2}}$(110) (Figure~\ref{fgr:Structures_a}) and \textbf{(b,d)} bridging O vacancies (O$_{\textit{br}}^{\textit{vac}}$) of reduced TiO$_{2-x}$(110), with $x$ = \sfrac{1}{8} (thin lines, Figure~\ref{fgr:Structures_b}) or \sfrac{1}{4} (thick lines, Figure~\ref{fgr:Structures_c}).  \textbf{(a,b)} $E_{\textit{ads}}$ calculated with PBE ($\medcirc$) and RPBE ($\meddiamond$) xc-functionals for (white) low (\sfrac{1}{4} and \sfrac{1}{2}ML), (turquoise) medium (\sfrac{3}{4} and 1ML), and (blue) high (1\sfrac{1}{4} and 1\sfrac{1}{2}ML) coverage. UPS difference spectra at \textbf{(c)} 150 K and \textbf{(d)} 300 K after 0.2 L exposure are from Ref.~\citenum{ThorntonH2ODissTiO2110}.  \textbf{(c,d)} Energies are relative to the VBM ($\varepsilon_{\textrm{VBM}}$).  Gray regions denote the clean surface DOS. Red dashed lines denote the highest PDOS peaks ($\epeak$) for 1ML H$_2$O@Ti$_{\textit{cus}}$ and \sfrac{1}{2}ML H$_2$O@O$_{\textit{br}}^{\textit{vac}}$. 
}\label{fgr:Spectra} 
\noindent{\color{JCTCGreen}{\rule{\textwidth}{1pt}}}
\end{figure*}

When a second layer of H$_2$O is added to the low coverage intact \sfrac{1}{2}ML H$_2$O@Ti$_{\textit{cus}}$ structure, the levels with H$_2$O 1b$_2$ character are unchanged, while the levels with 3a$_1$ and 1b$_1$ second layer character are more localized and weakly hybridized with the surface.  
These levels are seen as the two most intense peaks at $-4.3$ and $-2.2$~eV (Figure~\ref{fgr:Spectra}\textbf{(c)}).   The former coincides with the peak at $-4.2$~eV observed experimentally at low temperatures (Figure~\ref{fgr:Fig1}\textbf{(c)}), suggesting  multilayer H$_2$O structures may be present under these experimental conditions.  The intermolecular H bonding between the layers delocalizes the molecular levels of the first layer.  This is seen from the  peak at $-6.1$~eV with antibonding 3a$_1$ character on the first layer.  We saw the same behavior when increasing the first layer's coverage from \sfrac{1}{2}ML to 1 ML.  This is further confirmation that the peak observed experimentally at $-6.4$~eV has intermolecular character.

When a second \sfrac{1}{2} layer of H$_2$O is added to the 1ML H$_2$O@Ti$_{\textit{cus}}$ \{\sfrac{1}{3}D, \sfrac{2}{3}D\} structures \cite{LindanMultilayer}, a denser network of intermolecular and interfacial hydrogen bonds is formed, as shown in Figure~\ref{fgr:Structures_a}.  This causes a stronger hybridization between the OH and H$_2$O $\sigma$ levels.  For the \sfrac{1}{3}D structure, this results in the four distinct $\sigma$ peaks shown in Figure~\ref{fgr:Spectra}\textbf{(c)}.  On the one hand, the peaks at $-9.1$ and $-6.2$~eV have predominantly intact H$_2$O and O$_{\textit{cus}}$H character, as was the case for 1ML of \sfrac{1}{2}D H$_2$O@Ti$_{\textit{cus}}$. On the other hand, the peaks at $-7.9$ and $-7.4$ eV are most related to the second layer.  In effect, the H$_2$O $\sigma$ level of the second-layer H$_2$O, which is fully saturated with four hydrogen bonds, is upshifted by more than an eV.   

This is not the case for the \sfrac{2}{3}D structure (Figure~\ref{fgr:Structures_a}), where the peak at $-9.1$~eV instead has mostly intact second-layer H$_2$O 1b$_2$ character.  As was the case for intact 1\sfrac{1}{2}ML H$_2$O@Ti$_{\textit{cus}}$, the addition of a second \sfrac{1}{2} layer of H$_2$O induces a stronger hybridization of the O$_{\textit{br}}$H levels, and introduces an additional intense peak at $-4.4$~eV (Figure~\ref{fgr:Spectra}\textbf{(c)}).  This again suggests the experimentally observed peak at $-4.2$~eV (Figure~\ref{fgr:Fig1}\textbf{(c)}) may be due to multilayer H$_2$O.  

Overall, we find the addition of second-layer H$_2$O affects the resulting spectrum qualitatively.  We find both additional features and a redistribution of those due to the first H$_2$O layer.  When we instead add H$_2$O@Ti$_{\textit{cus}}$ to the \sfrac{1}{4}ML and \sfrac{1}{2}ML H$_2$O@O$_{\textit{br}}^{\textit{vac}}$ structures (Figures~\ref{fgr:Structures_b}, and \ref{fgr:Structures_c}) we find the resulting spectrum is the sum of the separate spectra to within 0.2 ~eV (Figure~\ref{fgr:Spectra}).  For example, the 1\sfrac{1}{2}ML \sfrac{1}{3}D spectrum (Figure~\ref{fgr:Spectra}d) for 1ML of intact H$_2$O added to \sfrac{1}{2}ML H$_2$O@O$_{\textit{br}}^{\textit{vac}}$ (Figure~\ref{fgr:Structures_c}) is basically the sum of the 1ML intact H$_2$O@Ti$_{\textit{cus}}$ (Figure~\ref{fgr:Fig1}\textbf{(a)}) and   \sfrac{1}{2}ML  H$_2$O@O$_{\textit{br}}^{\textit{vac}}$ (Figure~\ref{fgr:Fig2}\textbf{(a)}) PDOS spectra downshifted by 0.2~eV.  This explains the ease with which the experimental single-layer  H$_2$O spectra may be analyzed for levels outside the surface DOS region.

\subsection{Alignment of the Highest H$_{\text{2}}$O Occupied Levels}\label{Sect:Alignment}
So far, we have concentrated our analysis on the lower energy peaks observed in the experimental spectra. This was done to demonstrate the robustness of the calculated QP DOS.  Having established this, we now focus on the adsorbate levels near the VBM, which play an important role in photooxidation processes. In this respect, the highest H$_{\text{2}}$O occupied levels' alignment  for 1ML intact and dissociated H$_2$O@Ti$_{\textit{cus}}$,  and  \sfrac{1}{2}ML  dissociated  H$_2$O@O$_{\textit{br}}^{\textit{vac}}$ is of utmost importance. The former structure corresponds to the reactant species on stoichiometric surfaces\cite{MichaelidesDynamics}, which undergoes photo-irradiation. The latter structures act as hole traps and are thus the main oxidizing agents on TiO$_2$(110).\cite{PsalvadorJPCL2013,CR1995SurfaceBoudOHradicals}  

We have shown that the experimental peak at $-4.2$~eV\cite{ThorntonH2ODissTiO2110} is not, in fact, the highest energy peak of H$_2$O@Ti$_{\textit{cus}}$.  We instead find the highest-energy PDOS peak, $\epeak$, for 1ML intact H$_2$O@Ti$_{\textit{cus}}$ at $-1.1$~eV relative to the VBM (Figure~\ref{fgr:Spectra}\textbf{(c)}). This is 0.6~eV closer to the VBM than the $\sim-1.7$~eV estimate\cite{PSalvador2007} deduced from the onsets of the UPS difference spectra in   Ref.~\citenum{DeSegovia}. Moreover, as 1ML H$_2$O@Ti$_{\textit{cus}}$ dissociates,  $\epeak$ moves up to $-1.0$~eV (\sfrac{1}{2}D) and $-0.9$~eV (D) (Figure~\ref{fgr:Spectra}\textbf{(c)}).  This is again significantly higher than the $\sim-1.8$~eV estimate\cite{ImanishiJACS2007}  based on UPS difference spectra for the TiO$_2$(100) surface from Ref.~\citenum{Muryn1991747}\nocite{Muryn1991747}.  As was the case for CH$_3$OH on TiO$_2$(110)\cite{OurJACS}, this raising of $\epeak$ can be related to the charge transfer of $-0.4 e$ that accompanies deprotonation (arrows in Figure~\ref{fgr:Structures_a}).  We find for the 1ML intact structure on TiO$_2$(110) $\epeak$ is 0.2~eV closer to the VBM for H$_2$O than for CH$_3$OH\cite{OurJACS,MiganiLong,MiganiInvited}, while for the 1ML \sfrac{1}{2}D structures $\epeak$ is the same.  However, the highest PDOS peak is both less intense and broader for H$_2$O compared to CH$_3$OH, due to the stronger hybridization with the surface.  
This is why, as discussed in Section~\ref{Sect:Intact}, the QP $G_0W_0$ $\epeak$ is only $\sim0.1$~eV lower compared to DFT\cite{OurJACS,MiganiLong,MiganiInvited} (Figure~\ref{fgr:FigS1}).  After adding second-layer H$_2$O,  $\epeak$ is unchanged with weight mostly remaining on the first layer.

 We find for \sfrac{1}{2}ML  dissociated  H$_2$O@O$_{\textit{br}}^{\textit{vac}}$  $\epeak \approx-1.1$~eV relative to the VBM (Figure~\ref{fgr:Spectra}d), the same as for intact H$_2$O@Ti$_{\textit{cus}}$.  This is much higher than the previous estimate of $\sim-3.7$~eV\cite{ImanishiJACS2007} for O$_{\textit{br}}$H based on the UPS difference spectra in Ref.~\citenum{ThorntonH2ODissTiO2110}.  Our corrected $\epeak$ value agrees with the recently demonstrated photocatalytic importance of O$_{\textit{br}}$H sites as the main oxidizing species on TiO$_2$(110). \cite{PsalvadorJPCL2013} 

\old{From}\new{Based on} $\epeak$ for 1ML intact H$_2$O@Ti$_{\textit{cus}}$, \old{optical transitions}\new{vertical excitations} from the highest H$_2$O occupied levels to the TiO$_2$(110) conduction band require photon energies that  exceed the electronic band gap for bulk  rutile TiO$_{\text{2}}$ ($3.3 \pm0.5$~eV\cite{TiO2BandGap}) by $\gtrsim1$~eV. \old{Nevertheless, it can be argued that H$_2$O may not be \new{directly }photooxidized \old{as a result of}\new{via} these supra-band gap excitations. As a result,}\new{ However,} the hole \old{left behind}\new{generated} by \old{the electron}\new{such supra-band gap } excitation\new{s should be }\old{ will be }mostly located on TiO$_2$(110) O 2p$_{\pi}$  rather than  H$_2$O O 2p levels. This is because the H$_2$O  highest levels are hybridized with TiO$_2$(110) and are predominantly TiO$_2$(110) in character.

The fact that the highest H$_2$O levels are $\sim1$~eV below the VBM \new{does}\old{should} not \old{be used to justify their thermodynamically hindered photooxidation}\new{necessarily mean that they cannot be photooxidized} by holes photogenerated within the TiO$_2$(110) valence band.  A recent DFT study \old{using the hybrid HSE xc-functional has provided evidence for hole trapping}\new{with HSE06 found trapped holes} at  \old{HO--Ti$_{\textit{cus}}$}\new{ surface O sites, i.e., three-fold coordinated O$_{3\textit{fold}}$, are shared with nearyby H}O--Ti$_{\textit{cus}}$ \old{sites}\new{groups}.\cite{SprikSchematic} \new{Moreover, it has been suggested that H$_2$O can only be photooxidized, i.e., trap a hole, upon deprotonation\cite{JPCLDiValentin2013,SeloniPCETJACS2013}.  In other words, hole transfer to the HO--Ti$_{\textit{cus}}$ site should be mediated by the deprotonation of  intact H$_2$O@Ti$_{\textit{cus}}$ to the nearest O$_{\textit{br}}$ site.}  Altogether, this suggests that  H$_2$O@Ti$_{\textit{cus}}$ photooxidation should be initiated by band-to-band and supra-band photo-excitations, which result in the generation of holes within the TiO$_2$(110) valence band. These TiO$_2$(110) free holes may then be trapped at \old{TiO$_2$(110) surface O sites, i.e., three-fold coordinated }O\new{$_{3\textit{fold}}$ sites}, and \new{partially} transferred to nearby HO--Ti$_{\textit{cus}}$\old{ and O--Ti$_{\textit{cus}}$ sites}\new{ upon H$_2$O deprotonation}.\old{\cite{SprikSchematic} It has been suggested that this hole transfer to the HO--Ti$_{\textit{cus}}$ site is mediated by the deprotonation of  intact H$_2$O@Ti$_{\textit{cus}}$ to the nearest O$_{\textit{br}}$ site.\cite{JPCLDiValentin2013,SeloniPCETJACS2013}}

\subsection{\new{Vacuum Level Alignment}}\label{Sect:EvacAlignment}

\new{So far, we have considered the level alignment of the interfacial levels relative to the VBM of the substrate. This allows a direct comparison of the occupied PDOS with the measured UPS spectra.  However, to assess the photoelectrocatalytic activity of the interface, one needs the absolute level alignment relative to the vacuum level $E_{\textit{vac}}$. }

\begin{figure}[!t]
%\begin{figure}
%\noindent{\color{JCTCGreen}{\rule{\columnwidth}{1pt}}}
\includegraphics[width=\columnwidth]{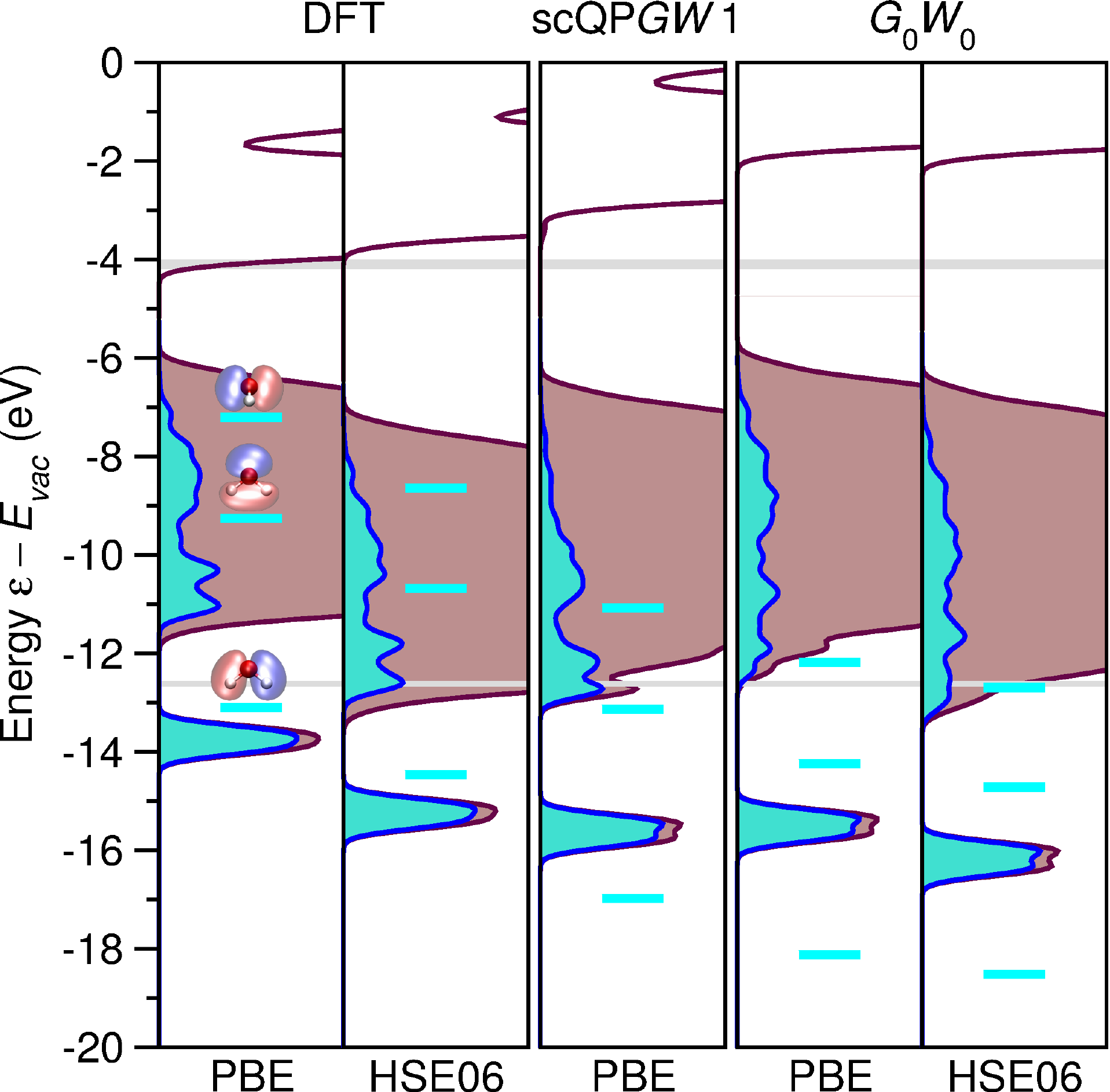}
\caption{\new{Absolute level alignment for 1ML intact H$_{\text{2}}$O adsorbed with antiparallel ($\rightleftarrows$) interfacial hydrogen bonds on coordinately unsaturated Ti sites (H$_2$O@Ti$_{\textit{cus}}$).  Total (maroon) and H$_2$O projected (blue) DOS computed with DFT, scQP$GW$1, and $G_0W_0$ using the generalized gradient approximation (PBE) and hybrid (HSE) xc-functionals.  Energies are relative to the vacuum level $E_{\textit{vac}}$. The measured $\varepsilon_{\textrm{CBM}}$ from Ref.~\citenum{SprikH2OAlignment} (thick gray line), measured and coupled-cluster (CCSD(T)) H$_2$O gas phase ionization potentials \textit{IP} from Ref.~\citenum{GWMoleculesMarquesJCTC2013} (thin gray line),  and for each level of theory the calculated gas phase 1b$_1$, 3a$_1$, and 1b$_2$ H$_2$O levels (marked in cyan) are provided.}}\label{fgr:EvacAlignment}
%\noindent{\color{JCTCGreen}{\rule{\columnwidth}{1pt}}}
\end{figure}
\begin{figure}[!t]
%\noindent{\color{JCTCGreen}{\rule{\columnwidth}{1pt}}}
\includegraphics[width=\columnwidth]{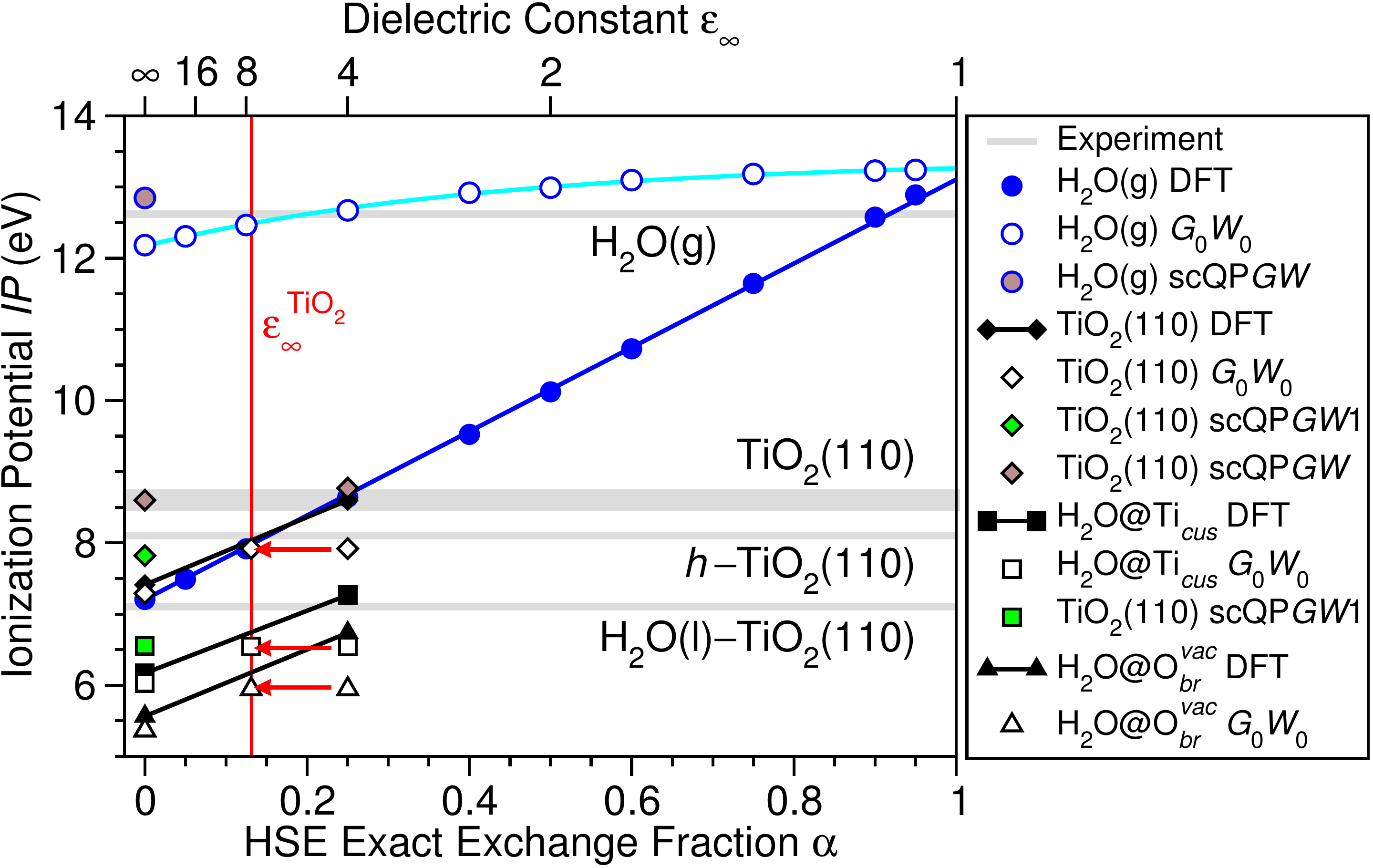}
\caption{\new{Ionization potential \textit{IP} versus exact exchange fraction $\alpha$ included in the HSE xc-functional and equivalent dielectric constant $\varepsilon_\infty \approx \alpha^{-1}$ from DFT (filled symbols), $G_0W_0$ (open symbols), scQP$GW1$ (green filled symbols), and scQP$GW$ (brown filled symbols) for H$_2$O in gas phase (circles), a clean \cite{MiganiLong} (diamonds) and a 1ML of intact H$_2$O@Ti$_{\textit{cus}}$ (squares) on the stoichiometric TiO$_2$(110) surface, and a \sfrac{1}{2}ML of dissociated H$_2$O@O$_{\textit{br}}^{\textit{vac}}$ (triangles) on the defective TiO$_{2-\text{\sfrac{1}{4}}}$(110) surface with \sfrac{1}{2}ML of O$_{\textit{br}}^{\textit{vac}}$.  The measured \textit{IP} for H$_2$O in gas phase\cite{GWMoleculesMarquesJCTC2013}, the stoichiometric TiO$_2$(110) surface \cite{MiganiLong,TiO2WorkFunction30,Onishi198833,C0CP02835E,Petek2PPH2O}, the $6-9$\% hydroxylated $h-$TiO$_2$(110) surface \cite{C0CP02835E}, and the liquid H$_2$O--TiO$_2$(110) interface\cite{SprikH2OAlignment} are shown in gray.  The self-consistent QP $GW$ \textit{IP} for H$_2$O in gas phase is indicated by the horizontal dashed line. The experimental dielectric constant of bulk TiO$_2$, $\varepsilon_\infty^{\text{TiO}_2} \approx 7.6$\cite{TiO2DielectricConstantExp}, averaged over the (110) surface is marked in red.  A linear fit to the DFT \textit{IP} (blue), and an exponential fit to the $G_0W_0$ \textit{IP} (cyan) for H$_2$O in gas phase are provided for comparison.}
}\label{fgr:SIFig1}
\noindent{\color{JCTCGreen}{\rule{\columnwidth}{1pt}}}
\end{figure}

\new{In Figure~\ref{fgr:EvacAlignment} we show the level alignment for gas phase H$_2$O and 1ML intact H$_2$O@Ti$_{\textit{cus}}$ relative to $E_{\textit{vac}}$ from DFT, scQP$GW$1, and $G_0W_0$ based on PBE and HSE xc-functionals.  These are compared to the measured CBM for the liquid H$_2$O--TiO$_2$(110) interface\cite{SprikH2OAlignment,GratzelNature2001}, and the measured and coupled-cluster (CCSD(T)) gas phase H$_2$O ionization potential \cite{GWMoleculesMarquesJCTC2013}. }

\new{ Our calculated \textit{IP} values for H$_2$O in gas phase are consistent with those reported previously in the literature\cite{GWMoleculesMarquesJCTC2013,GWMoleculesThygesenPRB2010,CarusoPRB2012,GWMoleculesEversJCTC2013}.  Although the relative energies of the 1b$_1$, 3a$_1$, and 1b$_2$ H$_2$O levels are consistent over all five levels of theory, the levels are rigidly downshifted. We observe a clear ordering in increasing \textit{IP}  of  PBE DFT (7.2 eV) $\less$ HSE DFT $\ll$ PBE~scQP$GW1$ $\less$ PBE~$G_0W_0$ $\lesssim$ HSE~$G_0W_0$ $\lesssim$ PBE~scQP$GW$ (12.8 eV) $\less$ Hartree Fock (HF 13.9 eV\cite{CarusoPRB2012}).}

\new{To understand the origin of this ordering, we have probed the dependence of the \textit{IP} on the fraction of}
\new{ Hartree-Fock}
\new{ exact exchange included in the range-separated HSE xc-functional via the parameter $\alpha$ in Figure~\ref{fgr:SIFig1}.  On the one hand, for DFT, we find a strong linear dependence of \textit{IP} on $\alpha$, i.e., $\textit{IP} \approx \textit{IP}_{\textrm{PBE}} + (\textit{IP}_{\alpha=1}-\textit{IP}_{\textrm{PBE}})\alpha \approx 7.2 +5.9\alpha$, with $\alpha \sim 0.9$ providing a quantitative agrement with experiment and CCD(T) calculations.  Overall, this linear dependence is not surprising, as $\alpha$ may be interpreted as the amount of electron-electron screening, i.e., the inverse dielectric constant $\varepsilon_\infty^{-1}$ \cite{Marques,GaliSelfConsistentEpsilonPRB2014}.  In other words, the fraction of exact exchange $\alpha$ included, determines the amount of screening, $\varepsilon_\infty^{-1}$, incorporated within the xc-functional.
The quantitative agreement of the \textit{IP} for $\alpha \sim 0.9$ is because small molecules, e.g., H$_2$O, are weakly screened in the gas phase ($\varepsilon_\infty \sim 1$).  

On the other hand, for $G_0W_0$, the calculated \textit{IP} has a much weaker dependence on $\alpha$, i.e., the starting xc-functional, with $\textit{IP} \approx \textit{IP}_{\alpha=1} - \Delta \textit{IP} (10^{\alpha-1}-1) \approx 13.4 - 1.2\times 10^{-\alpha}$.  Further, the $G_0W_0$ and DFT \textit{IP} coincide when $\alpha \rightarrow 1$.  For $G_0W_0$ based on PBE ($\alpha = 0$), the \textit{IP} already agrees semi-quantitatively with experiment, with full quantitative agreement obtained for $G_0W_0$ based on HSE06 ($\alpha = 0.25$).   This is because the RPA $\varepsilon_\infty \sim 1$, independently of $\alpha$.  Essentially, the calculated $G_0W_0$ \textit{IP}s  would also be obtained from DFT using an HSE xc-functional with $0.84 < \alpha < 1.0$, i.e., $ 1 < \varepsilon_\infty < 1.2$. Overall, this implies $G_0W_0$ is a predictive method for the \textit{IP} of small molecules.  However, the scQP$GW$ technique has the added advantage of being completely independent of the starting xc-functional \cite{CarusoPRB2012,MiganiLong}, while providing a nearly quantitative \textit{IP}. 
}

\new{For the H$_2$O--TiO$_2$(110) interface, e.g., 1ML intact H$_2$O@Ti$_{\textit{cus}}$,  the highest energy H$_2$O PDOS peak, $\epeak$, is pinned $\sim 1$~eV below the VBM across PBE DFT, HSE DFT, PBE sc$GW$1, PBE $G_0W_0$, and HSE $G_0W_0$.  For this reason, the \textit{IP} of the H$_2$O interfacial levels is controlled by the alignment of the VBM with respect to the vacuum.  This means we only need to consider the absolute VBM level alignment of the interface, i.e., the interface's \textit{IP} = $-\varepsilon_{\textrm{VBM}}+E_{\textit{vac}}$, as a descriptor of photoelectrocatalytic activity.}

\new{In Figure~\ref{fgr:EvacAlignment} we see that  the \textit{IP} of the interface follows a different ordering across the methodologies from that of gas phase H$_2$O.  In particular, we find PBE~$G_0W_0$ (6.0 eV) $\sim$ PBE DFT $\less$ HSE06~$G_0W_0$ $\approx$ PBE~scQP$GW$1 $\less$ HSE06 DFT (7.3 eV).   Figure~\ref{fgr:SIFig1} shows that, as was the case for H$_2$O in gas phase, the \textit{IP} of the H$_2$O@Ti$_{\textit{cus}}$ interface across the various methods is ordered according to the method's description of the screening, $\varepsilon_\infty^{-1}$.  

As discussed above, for hybrid xc-functionals such as HSE, the effective screening is determined by the fraction of exact exchange $\alpha$ included.  Essentially, $\alpha$ plays the role of the effective screening within the method, $\varepsilon_\infty^{-1}$.  Although HSE06 incorporates less screening ($\varepsilon_{\infty} \approx 4$) than experiment for rutile TiO$_2$ ($\varepsilon_\infty^{\textrm{TiO}_2} \approx 7.6$)\cite{TiO2DielectricConstantExp}, the HSE06 $IP$ for the interface is in agreement with the experimental estimate of $\textit{IP} \approx 7.1$~eV \cite{SprikH2OAlignment,GratzelNature2001}.    

If one performs $G_0W_0$ based on HSE06, a stronger screening is applied, i.e., $\varepsilon_\infty \approx 5.7$, yielding a lower \textit{IP} for the interface.  In fact, as indicated by the red arrow in Figure~\ref{fgr:SIFig1}, a similar \textit{IP} to HSE06 $G_0W_0$ should be obtained from HSE DFT by setting the fraction of exact exchange to the inverse dielectric constant of bulk TiO$_2$, i.e., $\alpha = 1/\varepsilon_\infty^{\textrm{TiO}_2}$.  Adjusting $\alpha$ to the measured inverse dielectric constant has been previously found to give improved band gaps\cite{Marques}.  From PBE scQP$GW$1, one obtains an \textit{IP} consistent with that of  HSE06 $G_0W_0$ .  This is because we find the screening in scQP$GW$ decreases from PBE RPA with each self-consistent cycle. Essentially, the final screening incorporated in scQP$GW$1 is similar to that of HSE06 RPA.

As shown in Figure~\ref{fgr:EvacAlignment}, PBE $G_0W_0$ gives an \textit{IP} slightly lower than PBE DFT for the interface, while the PBE $G_0W_0$ CBM is shifted up by about 2~eV.  This is surprising, since PBE DFT already yields a CBM level alignment for the interface in excellent agreement with experiment.  This is partially due to PBE RPA's overestimation of the screening of TiO$_2$ ($\varepsilon_\infty \sim 8.3$).  Although HSE06 $G_0W_0$ has a weaker screening than PBE $G_0W_0$, the resulting absolute alignment of the CBM is quite similar.  If instead, the self energy corrections are applied self-consistently via PBE scQP$GW$1, the absolute alignment of the CBM is significantly lower, but still greater than that of PBE DFT or HSE06 DFT.  This is again related to decreases in the dielectric constant with each self-consistent cycle.  For this reason, scQP$GW$1 tends to provide reasonable band gaps for TiO$_2$(110) interfaces.  Overall, we observe an ordering in increasing band gap of PBE DFT $\less$ HSE06 DFT $\lesssim$ PBE scQP$GW1$ $\less$ PBE $G_0W_0$ $\approx$ HSE06 $G_0W_0$, with HSE06 DFT providing the best absolute alignment of the CBM and VBM for the H$_2$O@Ti$_{\textit{cus}}$ interface.  }

\new{
In Figure ~\ref{fgr:SIFig1},  we show that a similar correlation between \textit{IP} and the method's description of screening holds for clean and hydroxylated $h-$TiO$_2$(110).  Specifically, we consider clean stoichiometric TiO$_2$(110) \cite{MiganiLong}, and dissociated H$_2$O@O$_{\textit{br}}^{\textit{vac}}$ on defective TiO$_{2-\text{\sfrac{1}{4}}}$(110) with \sfrac{1}{2}ML of O$_{\textit{br}}^{\textit{vac}}$.  Overall, $\textit{IP} \approx \textit{IP}_{\textrm{PBE}} + 5.9\alpha$ for all systems considered.  We again find that the \textit{IP} of PBE $G_0W_0$ $\sim$ PBE DFT, HSE06 $G_0W_0$ $\sim$ HSE($\alpha^{-1}=\varepsilon_\infty^{\textrm{TiO}_2}$) $\sim$ PBE $scGW1$, and PBE scQP$GW$ $\approx$ HSE06 scQP$GW$ $\sim$ HSE06 DFT.

HSE06 DFT provides the most accurate description of the \textit{IP} of the clean and H$_2$O@Ti$_{\textit{cus}}$ covered stoichiometric TiO$_2$(110) surfaces. Although the HSE06 DFT \textit{IP} for H$_2$O@O$_{\textit{br}}^{\textit{vac}}$ is significantly lower than the one measured for $h-$TiO$_2$(110), in both cases, the \textit{IP} is shifted to lower energies relative to the clean stoichiometric surface.  Differences in the magnitude of the shifts are probably due to the differences in defect coverage between the experiment (6--9\%)\cite{C0CP02835E} and the calculation (50\%). 

The similarty between HSE06 DFT and scQP$GW$ based on either PBE or HSE06 for the clean TiO$_2$(110) surface\cite{MiganiLong},  points to a similar screening from these two techniques. This also demonstrates the starting point independence of the scQP$GW$ technique.

To summarize,  although scQP$GW$  provides accurate \textit{IP}s, the band gap is greatly overestimated, as reported previously \cite{OurJACS,MiganiLong,KressescGW,PacchioniCatalLett2014}.  While scQP$GW1$ provides a more accurate band gap, it achieves only a qualitative description of the \textit{IP}.  
HSE06 achieves a quantitative description of both the \textit{IP} and band gap, but provides a poor description of the molecular level alignment relative to the VBM.\cite{OurJACS,MiganiLong,PacchioniCatalLett2014} However, since the highest occupied H$_2$O levels are significantly hybridized with the substrate, this is not a major drawback in this case.  
In general,  for TiO$_2$(110),  a more effective strategy is to combine the calculated \textit{IP} from HSE06 with the occupied interfacial levels' alignment from $G_0W_0$ or scQP$GW1$.} 

\section{CONCLUSIONS}\label{Sect:Conclusions}
 
The  level alignment prior to photo-irradiation is an important piece of the puzzle needed to get a complete atomistic picture of photocatalytic processes.   Here we have shown that the complex UPS spectra for the H$_2$O--TiO$_2$ interface may be disentangled using QP $G_0W_0$ PDOS.  \new{We have firmly established the robustness of the QP $G_0W_0$ H$_2$O PDOS by: (1) demonstrating its xc-functional (PBE, LDA, vdW-DF, and HSE06) independence, (2) comparing to self-consistent QP $GW$ techniques (scQP$GW1$), and (3) considering its dependence on surface coverage and dissociation.} Altogether, these calculations provide an accurate interpretation of the complex  UPS and MIES experiments\cite{DeSegovia,Krischok,ThorntonH2ODissTiO2110} for the H$_2$O--TiO$_2$(110) interface, and provide accurate estimates of the highest H$_2$O occupied levels' alignment relative to the VBM.

\new{Our results provide}\old{This provides} two important pieces\new{ of the puzzle}: (1) the molecular structure of the photocatalytic interface and (2) the molecular alignment of the doubly occupied levels near the VBM responsible for hole trapping prior to irradiation.  To complete the picture, the molecular structure and level alignment in the presence of the photo-generated hole is also needed.  Previous DFT studies using the hybrid HSE xc-functional have found a hole can be trapped at surface O 2p$_{\pi}$ levels of O$_{\textit{br}}$ and HO--Ti$_{\textit{cus}}$ sites \cite{SprikSchematic}.  However, the screening of such localized levels may not be well described by HSE, which tends to underbind localized interfacial levels \cite{MiganiLong}.  This underbinding is corrected upon inclusion of many-body effects via QP $G_0W_0$ \cite{MiganiLong}.   
Having demonstrated the capability of $G_0W_0$ for the description of level alignment prior to irradiation, this work points the way forward via future QP $G_0W_0$ studies of level alignment for trapped hole levels.

\old{Altogether, these calculations provide an accurate interpretation of the complex  UPS and MIES experiments\cite{DeSegovia,Krischok,ThorntonH2ODissTiO2110} for the H$_2$O--TiO$_2$(110) interface, and provide accurate estimates of the highest H$_2$O occupied levels' alignment relative to the VBM. }

\titleformat{\section}{\bfseries\sffamily\color{JCTCGreen}}{\thesection.~}{0pt}{\large$\blacksquare$\normalsize~}

\section*{ASSOCIATED CONTENT}
\subsubsection*{\includegraphics[height=8pt]{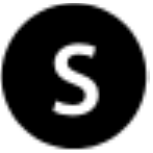} Supporting Information}
\noindent% Computational details, comparison of xc-functionals, O$_{\textit{br}}$ difference DOS, reduced TiO$_{2-\sfrac{1}{4}}$(110) structural schematics,  t
Total energies and optimized geometries.  This material is available free of charge via the Internet at http://pubs.acs.org.

\section*{AUTHOR INFORMATION}
\subsubsection*{Corresponding Author}
\noindent E-mail: annapaola.migani@cin2.es (A.M.)
\subsubsection*{Notes} 
\noindent The authors declare no competing financial interest.
\section*{ACKNOWLEDGMENTS} 
\new{We ackowledge fruitful discussions with Angel Rubio, }\old{W}and we thank S\old{.}\new{tefan} Krischok for providing experimental data.
We acknowledge funding from \old{the European Projects DYNamo (ERC-2010-AdG-267374) and CRONOS (280879-2 CRONOS CP-FP7); }Spanish Grants (FIS2012-37549-C05-02, \old{FIS2010-21282-C02-01, PIB2010US-00652,} RYC-2011-09582, JCI-2010-08156); Generalitat de Catalunya (2014SGR301, XRQTC); Grupos Consolidados UPV/EHU del Gobierno Vasco (IT-578-13); NSFC (21003113 and 21121003); MOST (2011CB921404); and NSF Grant CHE-1213189; and computational time from BSC Red Espanola de Supercomputacion and EMSL at PNNL by the DOE.

\bibliography{bibliography}

\end{document}